\documentclass[usenatbib,usegraphicx]{mn2e} 
\usepackage{amsmath,amssymb,amsfonts} 
\usepackage{graphics}         
\usepackage{color}          
\usepackage{alltt}

\usepackage{url}

\newcommand{\B}[1]{{#1}}
\newcommand{\AC}[1]{{#1}}

\newcommand{\ie}[0]{{i.e., }} 
\newcommand{\eg}[0]{{e.g., }}

\title{The SDSS Galaxy Angular Two-Point Correlation Function}

\author[Y. Wang et al.]{Y.~Wang$^1$, R.~J.~Brunner$^1$, and J.~C.~Dolence$^{1,2}$ \\
$^1$Department of Astronomy, University of Illinois, 1002 W. Green St., Urbana, IL 61801, USA\\
$^2$Department of Astrophysical Sciences, Peyton Hall, Princeton University, Princeton, NJ 08544 USA}

\begin{document}
\date{Submitted 2012 June}

\maketitle
\begin{abstract}
{
We present the galaxy two-point angular correlation function for galaxies selected from the seventh data release of the Sloan Digital Sky Survey. The galaxy sample was selected with $r$-band apparent magnitudes between $17$ and $21$; and we measure the correlation function for the full sample as well as for the four magnitude ranges: $17$--$18$, $18$--$19$, $19$--$20$, and $20$--$21$. We update the flag criteria to select a clean galaxy catalog and detail specific tests that we perform to characterize systematic effects, including the effects of seeing, Galactic extinction, and the overall survey uniformity. Notably, we find that optimally we can use observed regions with seeing $< 1\farcs5$, and  $r$-band extinction $< 0.13$ magnitudes, smaller than previously published results. Furthermore, we confirm that the uniformity of the SDSS photometry is minimally affected by the stripe geometry. We find that, overall, the two-point angular correlation function can be  described by a power law, $\omega(\theta) = A_\omega \theta^{(1-\gamma)}$ with $\gamma \simeq 1.72$, over the range $0\fdg005$--$10\degr$. We also find similar relationships for the four magnitude subsamples, but the amplitude within the same angular interval for the four subsamples is found to decrease with fainter magnitudes, in agreement with previous results. We find that the systematic signals are well below the galaxy angular correlation function for angles less than approximately $5\degr$, which limits the modeling of galaxy angular correlations on larger scales. Finally, we present our custom, highly parallelized two-point correlation code that we used in this analysis.}
\end{abstract}


\section{Introduction}

One of the most powerful and simplest probes of the galaxy distribution is the two-point angular correlation function, which quantifies the excess probability above a random distribution of finding one galaxy within a specified angle of another galaxy. For the case of a Gaussian random field, the two-point angular correlation function and its Legendre transform pair provide a complete statistical characterization of the galaxy clustering~\citep[see, \eg][]{Peebles}. Even for the case of non-Gaussianity, the two-point angular correlation function provides a simple and important statistical test of galaxy formation models~\citep{Teg04}.

\B{The two-point angular correlation function has been studied at bright magnitudes from the data releases from the Sloan Digital Sky Survey (SDSS) such as the Early Data Release~\citep[EDR;][]{Con}. This data release covered a few hundred square degrees of the in sky, and the two-point galaxy angular correlation function was calculated on scales from a few arc seconds to a few degrees. The measured correlation functions from the EDR were consistently found to obey a power law, $\omega(\theta) = A_\omega \theta^{(1-\gamma)}$, where $\gamma \simeq 1.7$ on small scales, with a break at $2^\circ$, beyond which the correlation dropped more steeply~\citep{Con}. For  deeper surveys, the power law relation of the small-scale correlation function held, with the amplitude decreasing at fainter magnitudes~\citep{Con}.}

\B{While these early SDSS results have provided a nice description of the angular clustering of galaxies, they only covered a relatively small area of the sky. In this paper, we present the measurement of the SDSS DR7 galaxy two-point angular correlation function. The SDSS DR7 galaxy sample covers nearly $10^4$ square degrees of the sky and includes approximately $10^8$ galaxies to a median redshift of $0.22$. Furthermore, in comparison to the SDSS EDR, the data processing techniques of the SDSS DR7 have been greatly improved~\citep{Aba04, Aba09}. The DR7 thus provides better image quality and photometric calibrations, with less severe systematic effects; and will, therefore, provide a more robust measurement of the galaxy angular clustering than previous large scale surveys.}

To accurately calculate the galaxy two-point angular correlation function, we must first minimize potential systematic effects in the galaxy catalog used to measure the correlation function. The systematics of the SDSS EDR were thoroughly studied by~\citet{Scr}. To minimize the systematic effects of seeing and Galactic extinction, they determined that the SDSS EDR galaxy sample had to be masked to exclude regions with seeing greater than $1\farcs75$ and reddening $ > 0.2$ magnitudes. 
Given the importance of minimizing the impact of systematic effects on the galaxy two-point angular correlation function and the significant changes that were made in the SDSS data processing pipeline between the SDSS EDR and the SDSS DR7, we have repeated many of the tests presented in~\citet{Scr} by using the SDSS DR7 data. In this paper we present the methods used to contain these systematic effects, the results of these systematic tests, the actual galaxy two-point angular correlation function for the SDSS DR7, and our massively parallel implementation that rapidly calculates correlation functions for large data sets.

In this paper, we first discuss the data and data samples in $\S$\ref{Data}, and we quantify the magnitude and source classification completeness limits in $\S$\ref{Complete}. After detailing our testing of the effects of different systematics and determining the optimal cuts to minimize their effects in $\S$\ref{SysResults}, we present the angular correlation function of galaxies and sub-samples split into magnitude bins in $\S$\ref{Correlation}. Next, we discuss our fast, tree-based correlation function code that we used to quickly calculate two-point angular correlation functions for these large data sets in $\S$\ref{Code}. Finally, we discuss these results and offer conclusions in $\S$\ref{Conclusion}.

\section{The Data}~\label{Data}

\B{The SDSS was a photometric and spectroscopic survey conducted by the Astrophysical Research Consortium at the Apache Point Observatory in New Mexico that was primarily designed to produce a data set to map large scale structure in the universe. The telescope was instrumented with either a wide-field, multi-band CCD camera or dual fiber-fed spectrographs. Cumulatively, the SDSS imaged over one-quarter of the entire sky, providing photometric information in five bands: $u$, $g$, $r$, $i$, and $z$~\citep{Fuku}. The data release studied herein, SDSS DR7, was released in November 2008, and includes objects observed through August 2008~\citep{Aba09}.}

\B{The main survey was centered on the north Galactic pole and was imaged in 37 interlaced stripes. Each stripe, which was observed during two days between the years $1999$--$2008$ is $2\fdg5$ wide, and the two ends of each stripe extend to low Galactic latitudes. The surveyed area includes a continuous portion in the northern Galactic hemisphere (34 stripes) and three individual stripes observed repeatedly in the southern Galactic hemisphere. In total, the data cover approximately $10^{4}$ deg$^{2}$ of the sky and consist of angular positions for around $10^{8}$ galaxies to a $5\sigma$ detection limit of $r \sim 23.1$~\citep{York}.}

\B{The photometric calibration was carried out by a separate 0.5-m photometric telescope adjacent to the SDSS main 2.5-m telescope (Photometric Telescope;~\citealt{Gunn}). A set of 157 standards stars, which covered the entire range in right ascension of the survey, were calibrated to the SDSS filter system~\citep{Smith}, and the main telescope observed these primary standards every night to quantify the relevant atmospheric extinction.}

\subsection{\em The Main Galaxy Sample}~\label{sample}

\begin{figure}
\vspace*{30pt}
\resizebox{8 cm}{!}{\includegraphics{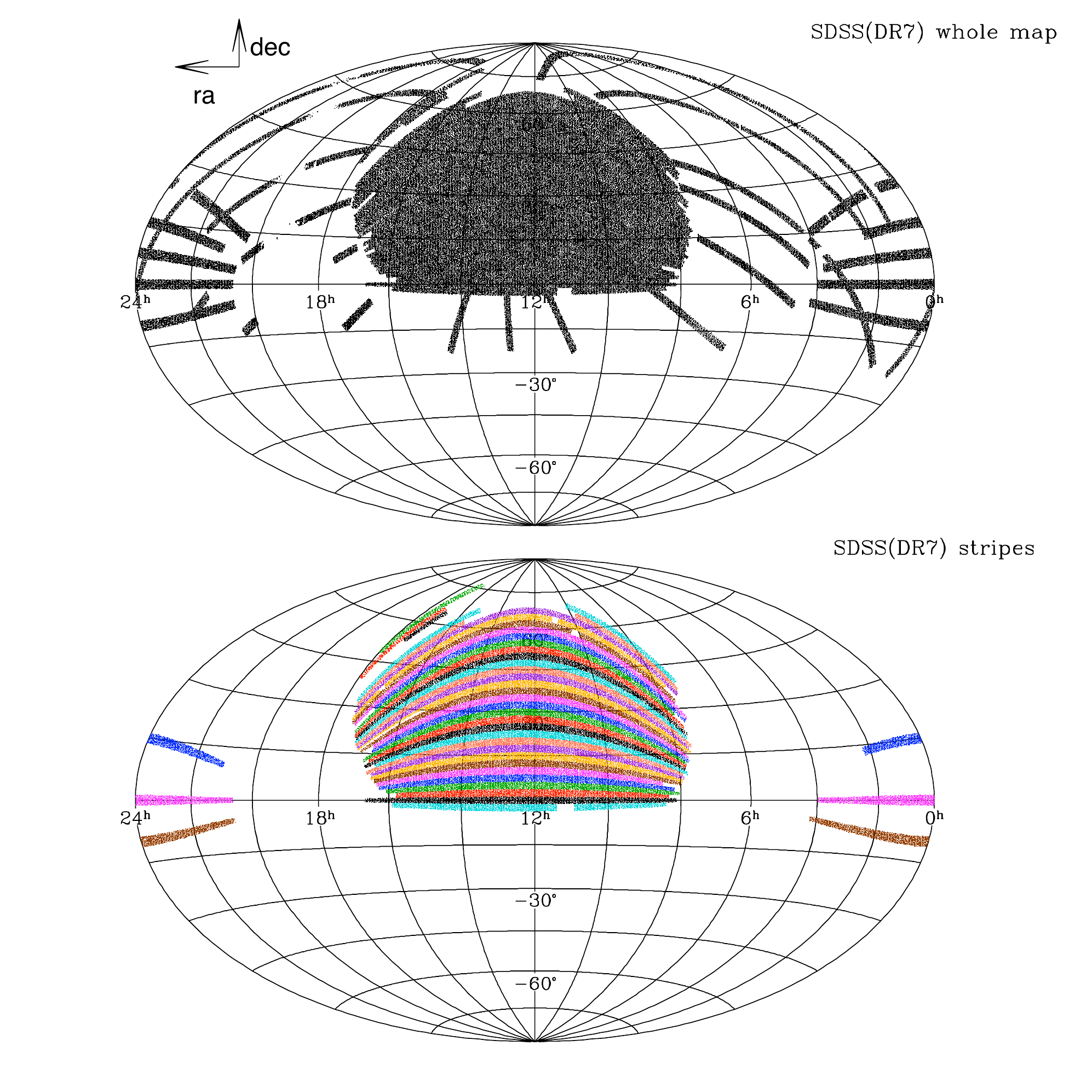}}
\caption{{Top: The full, primary data from the SDSS DR7. Bottom: The same data, but now showing only galaxies that are further cut to the theoretical SDSS footprint; restricted by observational flags and masked holes; and color-coded to indicate their SDSS stripe.}}
\label{skyDR7}
\end{figure}

The full data from the SDSS DR7 are shown in the top panel of Figure~\ref{skyDR7}, which contains galaxies and stars observed between March 1999 and August 2008. The complete procedure required to go from the SDSS data archive to our final galaxy sample is detailed in Appendix~\ref{AppCuts}; in this section we provide an overview of this process. Starting from the results of an SDSS CAS query that selected all objects with dereddened $g$, $r$, or $i$ magnitudes $< 23.0$, we first cut this sample to mask regions containing bright stars located within our Galaxy, and subsequently cut the remaining data to the theoretical footprint provided by the SDSS~\citep[\eg][]{Myers07}. Next, we restrict the sample to consist of all sources that pass the  appropriate flag tests as indicated by the SDSS project to select an observationally clean sample (the specific cuts used are described at \url{http://www.sdss.org/DR7/products/catalogs/flags.html} in the section entitled \textit{Clean sample of galaxies} and in Appendix~\ref{flagCut}) that consists of stars and galaxies.

\begin{figure}
\begin{center}
\resizebox{8 cm}{!}{\includegraphics{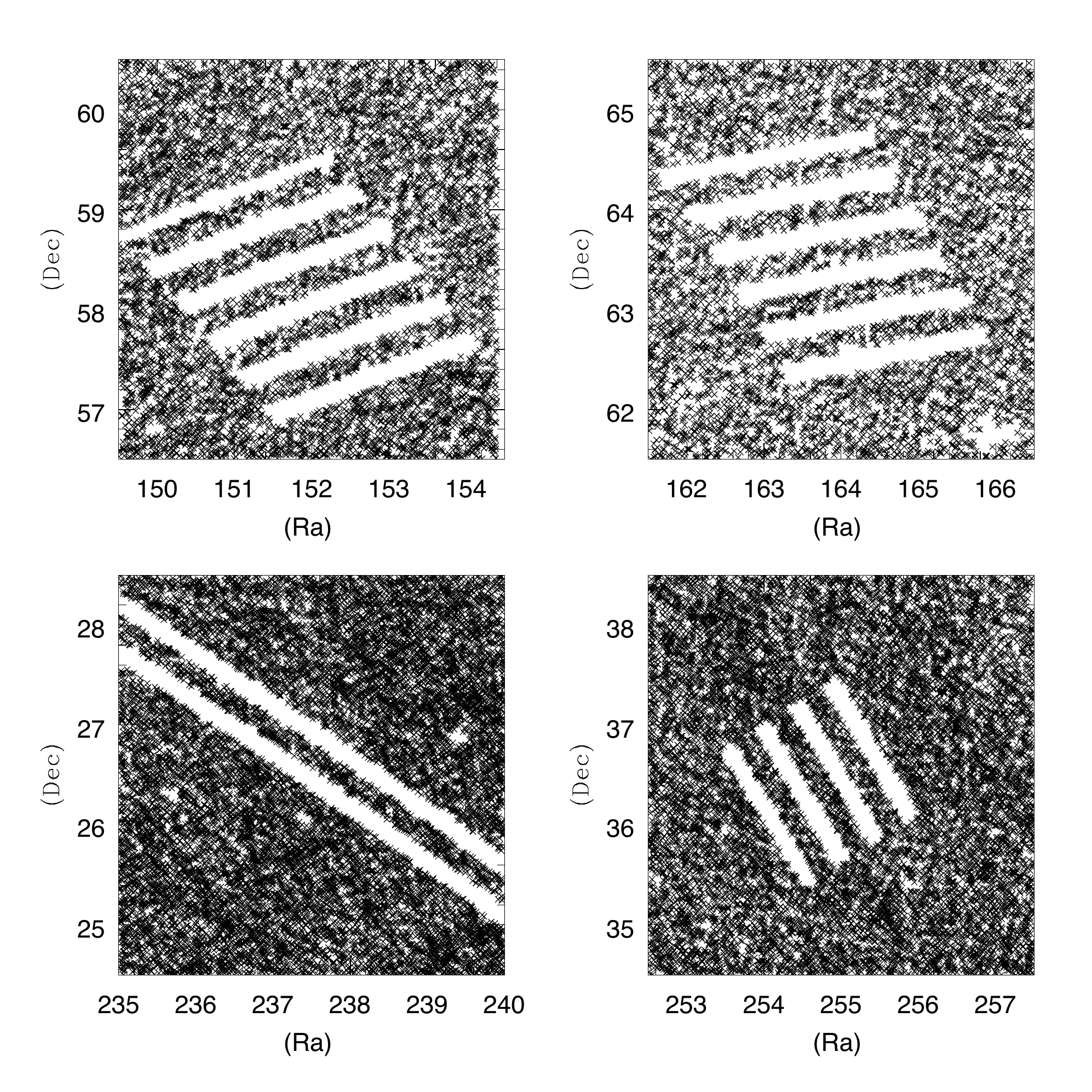}}
\end{center}
\caption{{Representative example areas in the SDSS DR7 footprint with missing data.}}
\label{holes}
\end{figure}

After restricting the data in the aforementioned manner, the data still include blank regions that lie within the survey area (see, \eg Figure~\ref{holes} for several examples). To simplify the process of masking these regions, we utilize the official survey $\lambda$/$\eta$ coordinates\footnote{\url{http://www.sdss.org/dr7/glossary/#survey_coords}} and manually check each stripe. Once an area of missing data is visually located, we identify the corners of the region bounding the missing data to an accuracy of $0.1$ degrees to further mask the affected region. As quantified in $\S$\ref{Complete}, we identify galaxies in this sample by using the SDSS \textit{type} parameter, and limit the entire sample to have extinction corrected $r$-band magnitudes within the range $17 < r \leq 21$, as specifically justified by the results presented in $\S$\ref{ML}. 

While the SDSS data set have been homogenized to the fullest extent possible, the data were observed in stripes that are each approximately $2\fdg5$ wide and of variable length (the stripes used in our angular correlation function analysis range from approximately $105\degr$ to $130\degr$ along the SDSS $\lambda$ coordinate). We select galaxies both from the northern Galactic hemisphere, which is a contiguous area of thirty-four stripes, and the southern Galactic hemisphere, which  has only three stripes. In the bottom panel of Figure~\ref{skyDR7}, we present our final galaxy sample, color-coded to indicate the SDSS stripe to which they belong.

In the end, our data cover $\sim8$,$000$ deg$^2$ of the sky. The final galaxy sample we analyze (\ie $17 < r \le 21$) contains nearly $22$ million galaxies with a median redshift of $z = 0.21$. To quantify the dependence on magnitude of our galaxy two-point angular correlation measurements, we split the full galaxy sample by magnitude into four sub-samples: $17<r\leq18$ ($\sim0.8$ million galaxies), $18<r\leq19$ ($\sim2.5$ million galaxies), $19< r \leq20$ ($\sim7.2$ million galaxies), and $20 < r \leq 21$ ($\sim19.3$ million galaxies). 

\subsection{\em Stripe 82 Coadd Data}~\label{coadd}

While the SDSS data have been carefully reduced and calibrated, we still need to quantify the limiting magnitude of the main sample for  cosmological analyses. To identify this magnitude limit, we need to compare the SDSS data to a deeper, more complete data set over as wide an area as possible. While several options exist for making this comparison, in the end we selected to use the coadded Stripe 82 data produced by the SDSS Legacy Survey that were also published as part of the SDSS DR7~\citep{Aba09}. While not as deep as other possible data sets, these data have the benefit of being taken with the same telescope and instrument as the main SDSS DR7 photometric data. And, after the coaddition of the individual observations, these data were reduced with the same data processing software stack~\citep{Annis}, thereby minimizing any systematic differences between the main and test data sets.

The SDSS Legacy Survey was a 3-year extension of the original SDSS that began operations in July 2005 and completed in July 2008. This legacy survey contains data  from both the SDSS-I and SDSS-II projects, and covers more than $7$,$500$ square degrees of the northern Galactic hemisphere and $740$ square degrees of the southern Galactic hemisphere. One of the primary science drivers for the SDSS-II project was to detect and measure light curves for a large number of supernovae~\citep{Frieman}. As a result, the SDSS southern equatorial stripe 82  was repeatedly imaged during this survey extension during the months of September, October, and November (\ie the three months when this stripe could be observed at the lowest airmass) in each of the three years: 2005--2007~\citep{Aba09}. In the interest of constructing dense light curves for variable supernovae, these photometric data were acquired even when conditions were non-optimal. 

The SDSS has released 123 runs that cover the Stripe 82 footprint\footnote{\url{http://www.sdss.org/dr7/coverage/sndr7.html}}, which have been observed under variable seeing, sky brightness, and photometric conditions. The best runs have been coadded by the SDSS collaboration to produce a final Stripe 82 coadded catalog, in which any given region has been observed between $20$ and $40$ times. Thus, the final Stripe 82 coadded catalog is nearly two magnitudes deeper than a single SDSS observation~\citep{Annis}, and covers an area $2\fdg5$ wide and $\sim110 \degr$ long, ranging from  $-50\degr$ to $60\degr$ in right ascension (as this is an equatorial stripe, right ascension is approximately equivalent to $\lambda$, which is the survey longitude coordinate). As a result, we use these coadded Stripe 82 data to define the completeness limits of the main DR7 sample, which is discussed in Section~\ref{Complete}.

We selected the deeper, coadded data covering the Stripe 82 footprint by following the same procedures used for the main galaxy sample, but now applied to the SDSS CAS Stripe 82 Catalog\footnote{\url{http://cas.sdss.org/stripe82/}}. Specifically, we first use the same query specified in Appendix~\ref{query} to select the Stripe 82 coadded data, after which we cut these data to the Stripe 82 footprint  as described in Appendix~\ref{footprint}, and we finally select clean detections by employing the flag cuts as described in Appendix~\ref{flagCut}. This produces a sample of $\sim$8.4 million sources from the Stripe 82 coadded data (hereafter `coadd'). In the same manner, we also select sources (both galaxies and stars) from the full DR7 catalog that lie within the Stripe 82 footprint (hereafter `main sample'), which consists of $\sim$4.3 million sources.

\section{Completeness Limits}\label{Complete}

When making cosmological measurements from the full SDSS DR7 data, we wish to be as inclusive as possible while minimizing any systematic effects. By using the SDSS EDR data, which were denoted by starred magnitudes (\eg $r^*$) as opposed to the final unstarred magnitudes (\eg $r$), \citet{Scr} suggested that $r^* \la 22$ was sufficient. A later analysis of the SDSS EDR data by~\citet{Inf}, however, suggested a brighter limit of $r^* \leq 20.5$ was more appropriate. In addition, a subsequent SDSS analysis demonstrated that the photometric pipeline used to process the SDSS EDR data incorrectly produced a $0.2$ magnitude offset~\citep{Aba04}, which was corrected in later data releases. As a result, before addressing any other specific systematic effects,  we must first identify the magnitude range over which large-scale photometric analyses can be reliably performed with the SDSS DR7 data. This requires that we cross-match the main sample data to the deeper, coadd data within the Stripe 82 footprint.

\begin{table}
\begin{center}
\caption{The percentage of matched sources between the Stripe 82 main sample and the Stripe 82 coadd data, split into all galaxies, all stars, and galaxies and stars in the magnitude range $17 < r \le 21$.}
{\begin{tabular}{c c  c c c}\\ \hline \hline
$r$-band model & Galaxies & Stars & Galaxies & Stars\\
magnitude difference & &  & \multicolumn{2}{c}{$17 <r\leq 21$}\\ \hline
$0.1$ & 42.3\% & 73.8\% & 63.3\% & 95.1\% \\
$0.2$ & 64.0\% & 88.7\% & 81.2\%& 98.8\% \\
$0.5$ & 89.9\% & 98.7\% & 93.8\% & 99.7\% \\
$1.0$ & 98.2\% & 99.8\% & 97.9\% & 99.8\% \\ \hline
\label{match}
\end{tabular}}
\end{center}
\end{table}

\subsection{\em Cross-Matching Between Catalogs}

When matching sources between two surveys, there are typically two restrictions that can be used to correctly identify the same source in both surveys. The first restriction is the use of a distance limit to force matched sources to be physically close on the sky, while the second restriction is a magnitude limit that forces matched sources to have similar measured fluxes. In our case, we are matching between two surveys that use the same telescope, imaging camera and data reduction pipeline, with the only real difference being that the coadd data are measured from an image that results from the combination of a large number of observations taken in varying conditions over a number of different years. Thus we felt that while our matching algorithm must employ a small distance tolerance for a successful match, we did not feel a magnitude restriction was appropriate.

As a result, to match objects between the main and the coadd samples, we only imposed a distance limit of $0\farcs56$, which is the approximate diagonal size of an SDSS camera pixel. Once the matching between the two surveys was completed, we calculated the differences in the dereddened $r$-band model magnitudes between the matched sources, and tabulate the results in Table~\ref{match}. Overall, approximately \B{56.4\%} of the matched objects have a magnitude difference less than $0.1$, and about \B{75.1\%} have a magnitude difference less than $0.2$, although it is also clear that galaxies show considerably larger magnitude differences than their stellar counterparts. 

After exploring this issue in more detail, primarily by visually inspecting a number of matched sources with large magnitude differences, we have found three primary reasons for the relatively high number of sources with larger than expected magnitude differences. First, the observations used to construct the coadd image were taken over a number of years, allowing for source photometric variability to induce magnitude differences. Second, the coadd image extends fainter than a standard, single pass SDSS image, and will, therefore, have a lower background sky level. This means that the SDSS processing pipeline will probe to a lower surface brightness, which can result in a change in the measured size of a galaxy and thus its model magnitude. Finally, the deeper coadd image will also contain more sources, which will lead to crowding issues that can complicate both source deblending and pixel assignment. These will also both change the measured size of a galaxy and thus its model magnitude. As a result of these effects, we feel confident in the use of this cross-matched catalog to determine a suitable magnitude limit for our main sample data.

\begin{figure*}
\begin{center}
\resizebox{8 cm}{!}{\includegraphics{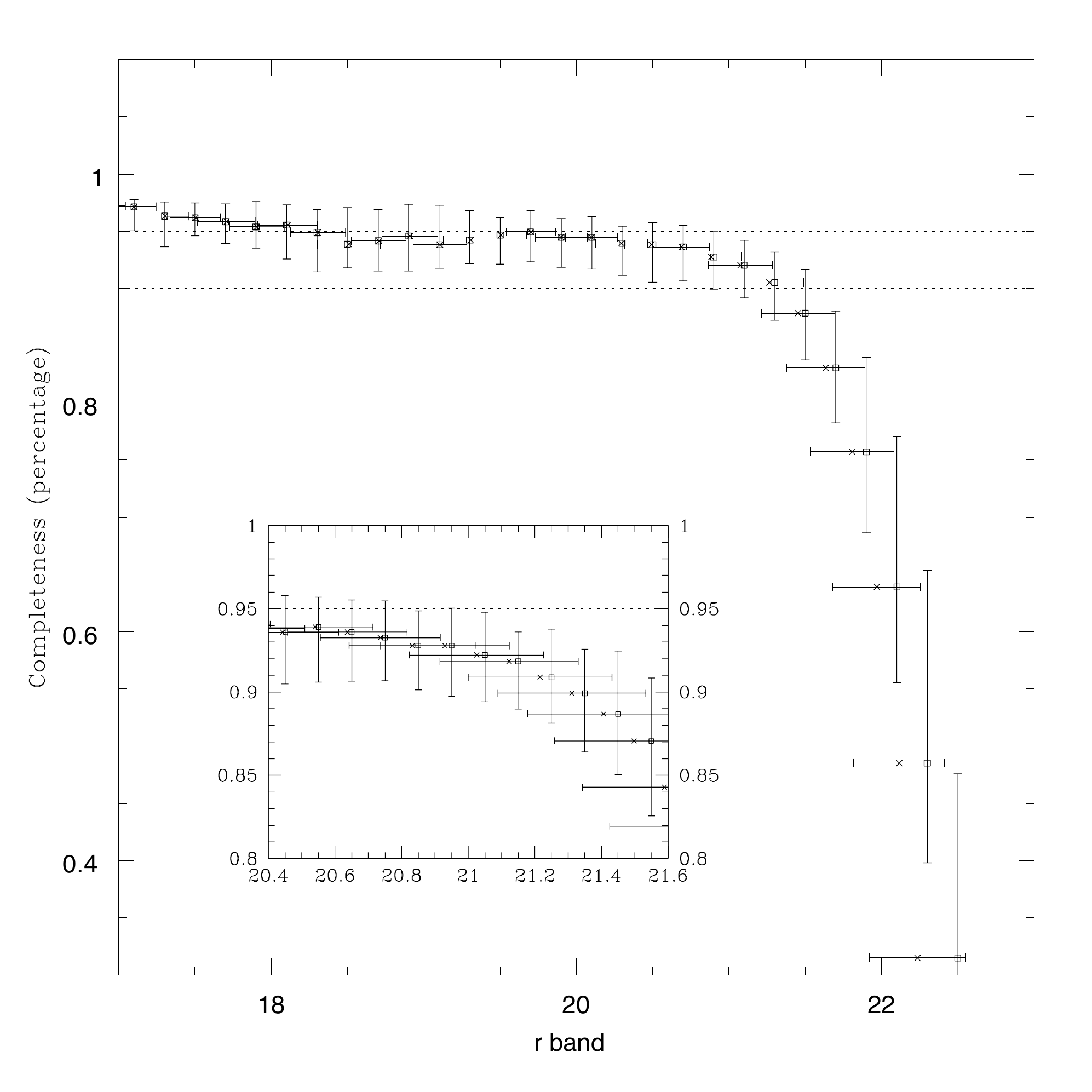}}
\resizebox{8 cm}{!}{\includegraphics{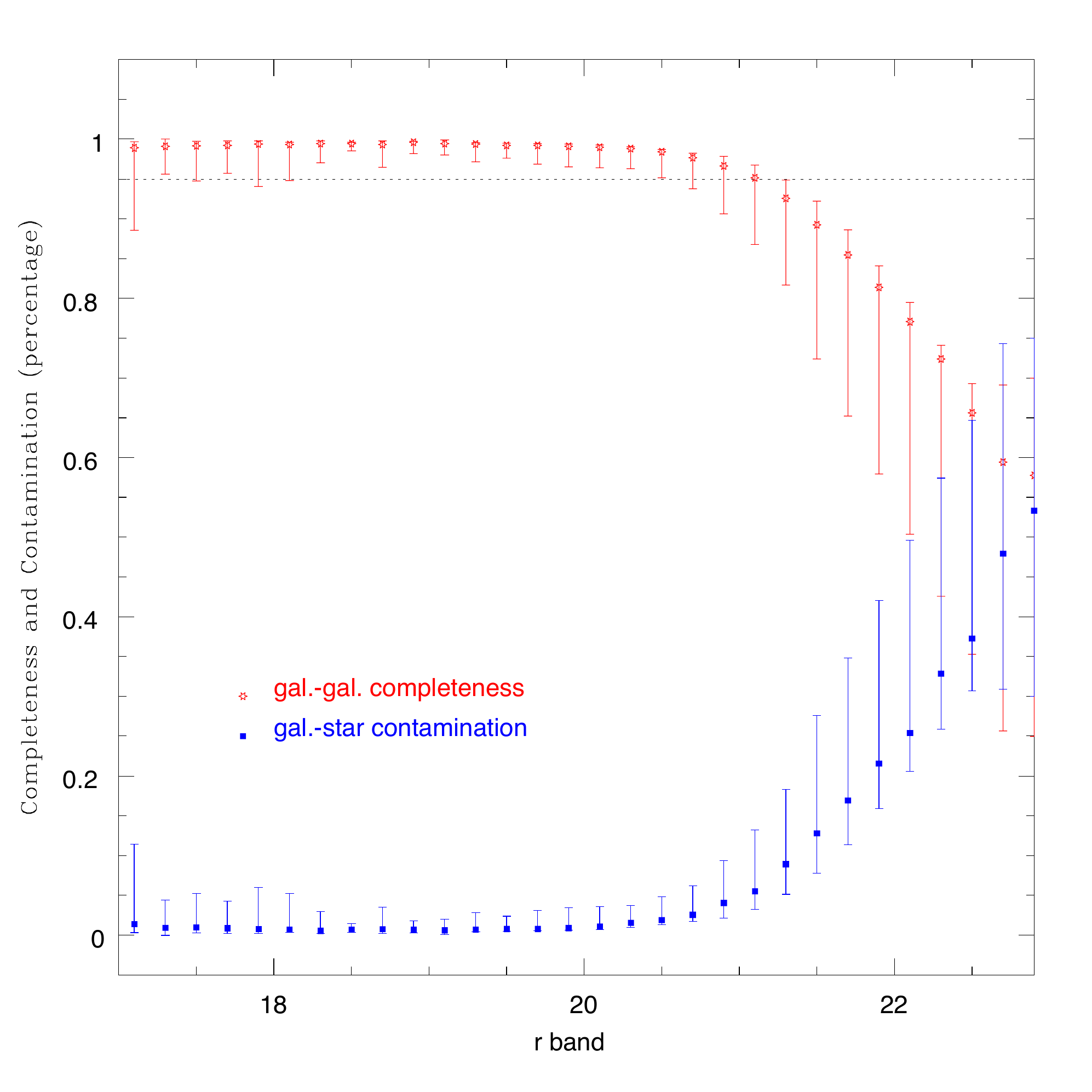}}
\end{center}
\caption{{Left: The detection completeness of sources in the main sample as a function of their dereddened $r$-band model magnitude. The squares and vertical error bars show the median, minimum, and maximum fraction of matched sources between the coadd and main sample as a function of the coadd magnitude. The crosses and horizontal error bars are the mean and standard deviation of the main sample magnitudes for the matched sources, showing that the match fraction remains above 90\% complete to $r \sim 21$ for the main sample. Right: The classification completeness and contamination of main sample galaxies as a function of their dereddened $r$-band model magnitude, showing that we are above 95\% complete at $r = 21$. The completeness (contamination) is measured by identifying galaxies (stars) in the deeper, coadd data that are classified as galaxies in the main sample. The points indicate the median value, while the upper and lower limits correspond to the maximum and minimum values, respectively.}}
\label{compltfig}
\end{figure*}

\subsection{\em Magnitude Limit}\label{ML}

After constructing the cross-matched catalog, we first look to identify the magnitude limit we must impose on the main sample data. To do this, we use the deeper coadd data as a guide to indicate where the main sample becomes incomplete. To quantity this limit, we divide the Stripe 82 footprint into \B{$10$} chunks. Within each of these chunks, we compute the fraction of sources in the coadd data that are matched to sources in the main sample in bins of width $0.2$ magnitudes. Since not all coadd data are matched, this process begins by using the coadd $r$-band, dereddened model magnitudes. 

By combining the matched fractions within a given magnitude bin across all chunks, we obtain a distribution that characterizes the detection completeness of the main sample as a function of the coadd $r$-band magnitude. We present the minimum, maximum, and median values of these distributions as the vertical error bars and square points, respectively, in the left plot of Figure~\ref{compltfig}. From this distribution, we see that the median value remains consistent with 90\% completeness or better to a dereddened, coadd $r$-band model magnitude limit of $r = 21$. 

However, since we must apply this magnitude limit to the entire SDSS DR7 main galaxy sample, we need to compute the corresponding dereddened $r$-band model magnitude limit for the main sample. To do this, we take the distribution of matched sources across all chunks within a given coadd magnitude bin, and compute the mean and standard deviation of the main sample magnitudes for all sources (we do exclude all non-detections from the main sample in this calculation). We present these values as the crosses and horizontal error bars in the left plot of Figure~\ref{compltfig}, which indicates that the same magnitude limit of $r \sim 21$ is appropriate for the main sample. We further confirmed this result by verifying that the average difference between the dereddened $r$-band model magnitude for a main sample source and the same source in the coadd is consistent with zero, with an increasing dispersion to fainter magnitudes as expected.

\subsection{\em Star/Galaxy Classification}\label{SG}

The detection completeness is only one part of the picture, however, as we also must know the accuracy of the SDSS pipeline's source classification as a function of  dereddened $r$-band model magnitude. To compute the classification completeness, we repeat the analysis in the previous section, but now start with sources classified in the main sample as galaxies (\ie $type = 3$). Specifically, we compute the fraction, within each chunk in bins of width $0.2$ magnitudes, the fraction of main sample galaxies classified as galaxies in the deeper, coadd data (\ie the classification completeness) and as stars in the deeper, coadd data (\ie the classification contamination). 

From these distributions, we compute the minimum, maximum, and median fractional values as a function of the main sample dereddened $r$-band model magnitude. We present these results in the right-hand  panel of Figure~\ref{compltfig}, where the galaxy completeness is displayed in red and the stellar contamination is displayed in blue. In either case, the minimum and maximum fractional values are displayed as the error bars while the median values are shown as the points. From this figure, we see that our completeness is above 95\% at our previously stated dereddened $r$-band model magnitude limit of $r = 21$, and in fact that source classification is reliable over the entire magnitude range of $17 < r \leq 21$.

\section{Results of Systematic Tests from DR7}\label{SysResults}

\citet{Scr} performed a detailed analysis by using the SDSS EDR to quantify possible systematic effects on clustering measurements that use the SDSS main galaxy sample. This work was leveraged repeatedly by subsequent authors, including to measure the galaxy two-point angular correlation function~\citep{Con} and the galaxy angular power spectrum~\citep{Tegmark02}. With later SDSS data releases, new constraints for either Galactic extinction or seeing were adopted, as predicated by a correlation function~\citep{R06} or an angular power spectrum~\citep{Hayes12}. More recently, ~\citet{R11} have performed a detailed analysis of the effects of systematics in the SDSS DR8 on the clustering of luminous red galaxies, in particular finding that stars have become more problematic in this newer data release. As a result, in this section we perform a detailed study of different systematics effects in the SDSS DR7 main galaxy sample. We note that all of these tests are done in two-dimensions, and can, therefore, be applied to any angular measurement of a two-dimensional survey data set.

\subsection{\em Pixelisation}\label{pixelisation}

In order to quantify certain discrete systematic effects, we must sample the galaxy distribution on similar physical scales as the relevant systematic effects. To accomplish this, we divide the relevant data into small pixels, or cells, and measure the fluctuations of a particular systematic effect (\eg seeing or reddening) across this distribution of pixels. Given the distinct scanning strategy of the SDSS survey, a specialized, pseudo-rectangular, approximately equal-area pixelisation strategy was developed by Tegmark, Xu, and Scranton (SDSSPix\footnote{\url{http://dls.physics.ucdavis.edu/~scranton/SDSSPix/}}) that works in SDSS $\lambda/\eta$ coordinates~\citep{Sto}. As a result, we use SDSSPix to quantify the density of sources within the SDSS stripe-based geometry for all relevant systematic tests.

SDSSPix has been used to measure the correlation function for a pixelised SDSS sample~\citep[see, \eg][]{Scr,R06}, but \citet{Hayes12} demonstrated that SDSSPix can bias a clustering measurement since the pixels are not the same size across a given stripe (the ratio of the pixel height to the pixel width decreases towards the ends of a stripe). As a result, we follow~\citet{Hayes12} and explore the use of a second pixelization scheme, HEALPix, to compute our pixelised correlation functions. HEALPix was developed by~\citep{Gorski05} and works in any spherical coordinate system. HealPix creates 12 equal-area curvilinearly base-patches,  from which pixels are generated at higher resolutions with either a RING or NESTED numbering scheme. 

\begin{figure}
\begin{center}
\resizebox{15 cm}{!}{\includegraphics{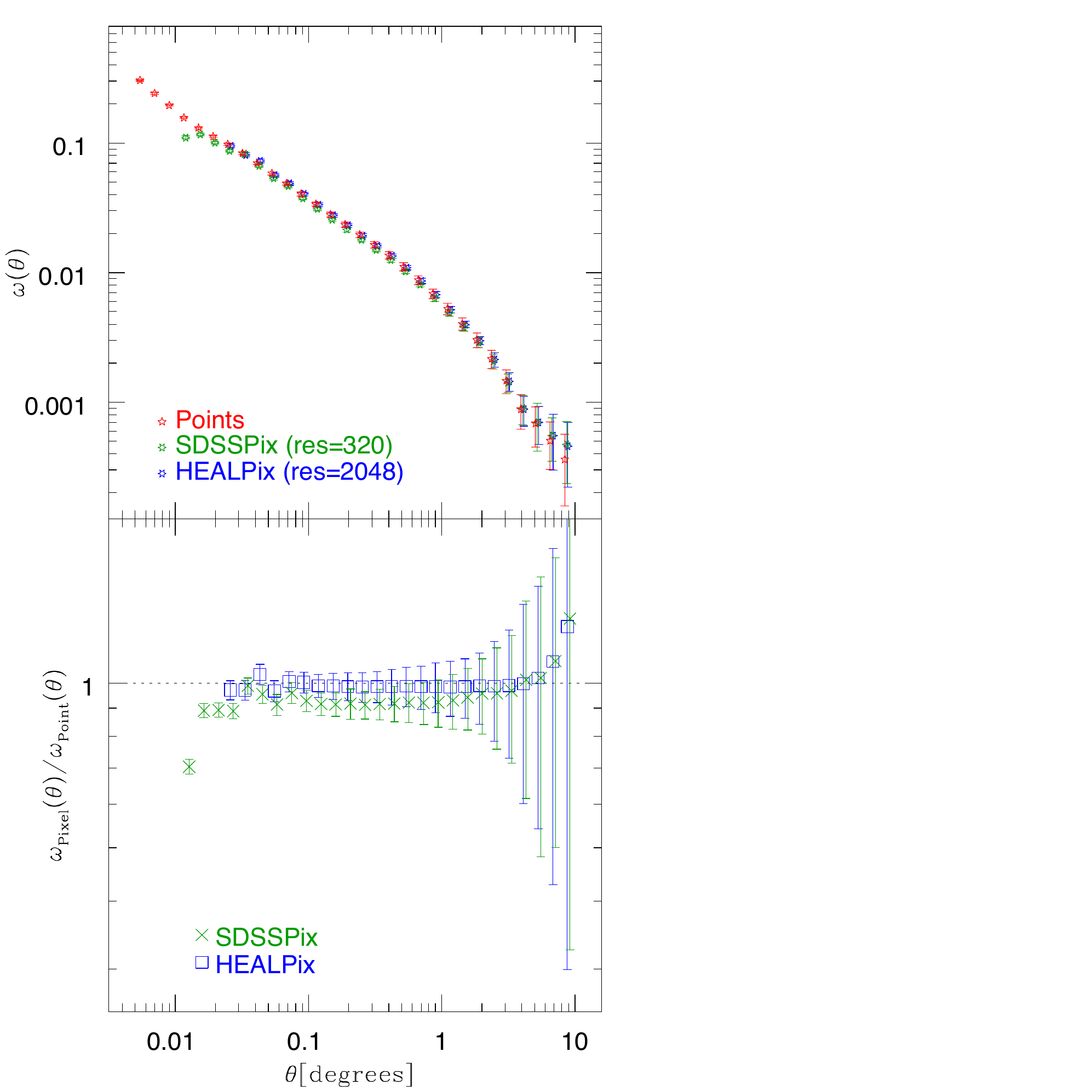}}
\end{center}
\caption{{Top: A comparison between the pixel-based (HEALPix resolution 2048 and SDSSPix resolution 320) and the point-to-point based pair count methods used in this paper. Bottom: The ratio of the above pixel-based correlation to the point-to-point based correlation. The errors are calculated by propagation of jackknife errors in quadrature.}}
\label{estiComPix}
\end{figure}

To decide which pixelisation scheme is optimal for our systematic tests, we pixelate the SDSS DR7 with both schemes, using SDSSPix at resolution 320 and HEALPix at resolution 2048 \B{(these resolutions produce equal area pixels: 3.10 square arcminutes for SDSSPix and 2.95 square arcmintutes for HEALPix)}. We compute the two-point angular correlation function for the SDSS DR7 data by using the point-to-point method described in $\S$\ref{Code} and the pixel based method described in $\S$\ref{CrossFunMethod}. The results from all three methods are directly compared in the top panel of Figure~\ref{estiComPix}, while the bottom panel compares the ratio of the pixel based methods to the point-to-point method. From this figure, in particular the ratio plot in the bottom panel, we see that SDSSPix systematically underestimates the correlation function, which becomes more severe at smaller angles (we believe this is a manifestation of the changing pixel shape). As a result, we adopt the HEALPix scheme for all pixel based systematic correlation function tests.

\subsection{\em Density Fluctuations Among Stripes}\label{numDenMethod}

The SDSS survey observed data along great circles, which are known as stripes that are identified by their stripe number. To explore the effects of this observing strategy on the uniformity of the full galaxy sample, we examined the uniformity of the galaxy counts, including as a function of magnitude, across these different stripes. For this test, we first used the SDSS algorithm to cut the full sample into the thirty-seven  constituent stripes present in the SDSS DR7 data\footnote{\url{http://cas.sdss.org/dr7/en/help/docs/algorithm.asp?key=resolve}}.

\begin{figure*}
\begin{center}
\resizebox{8 cm}{!}{\includegraphics{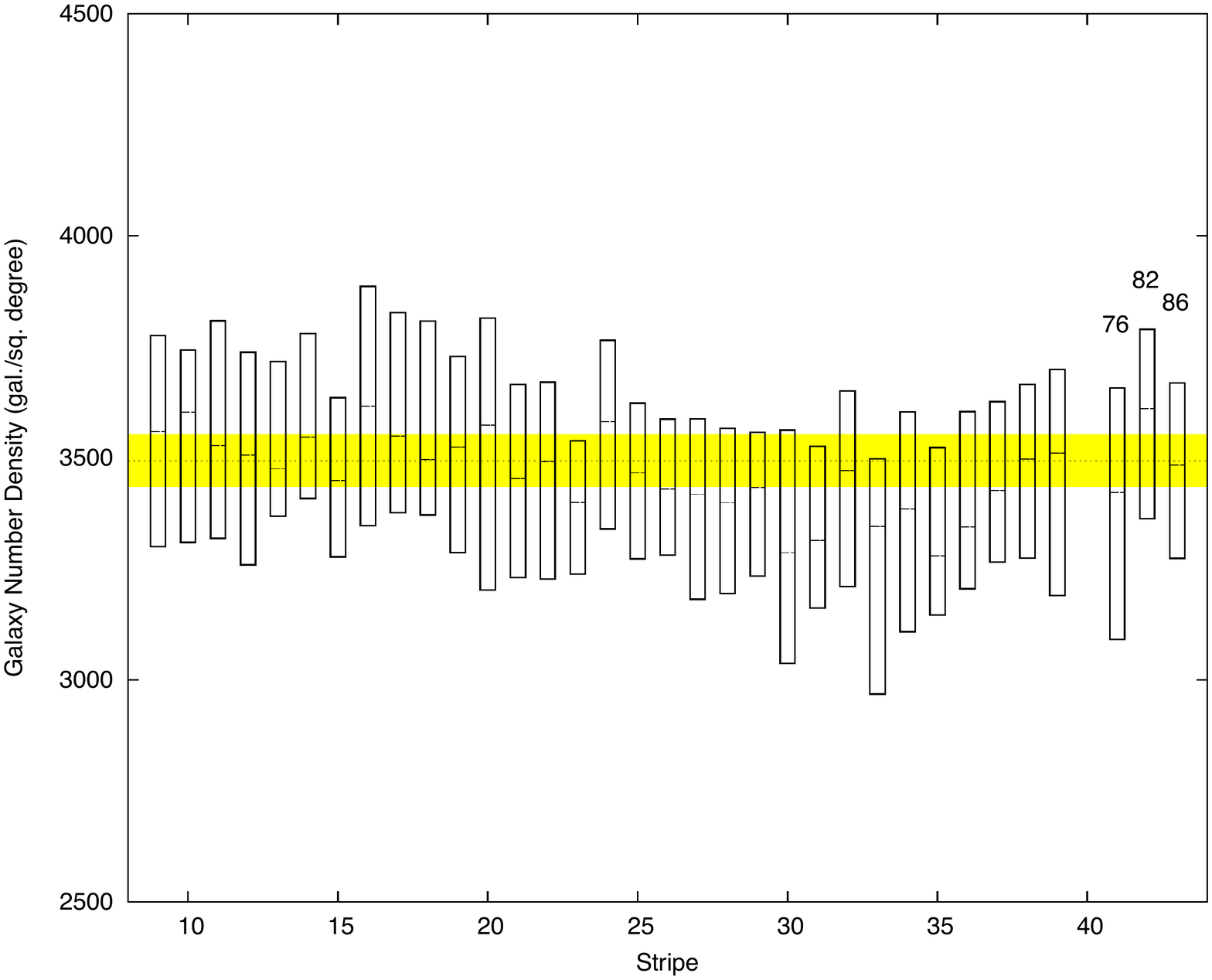}}
\resizebox{8 cm}{!}{\includegraphics{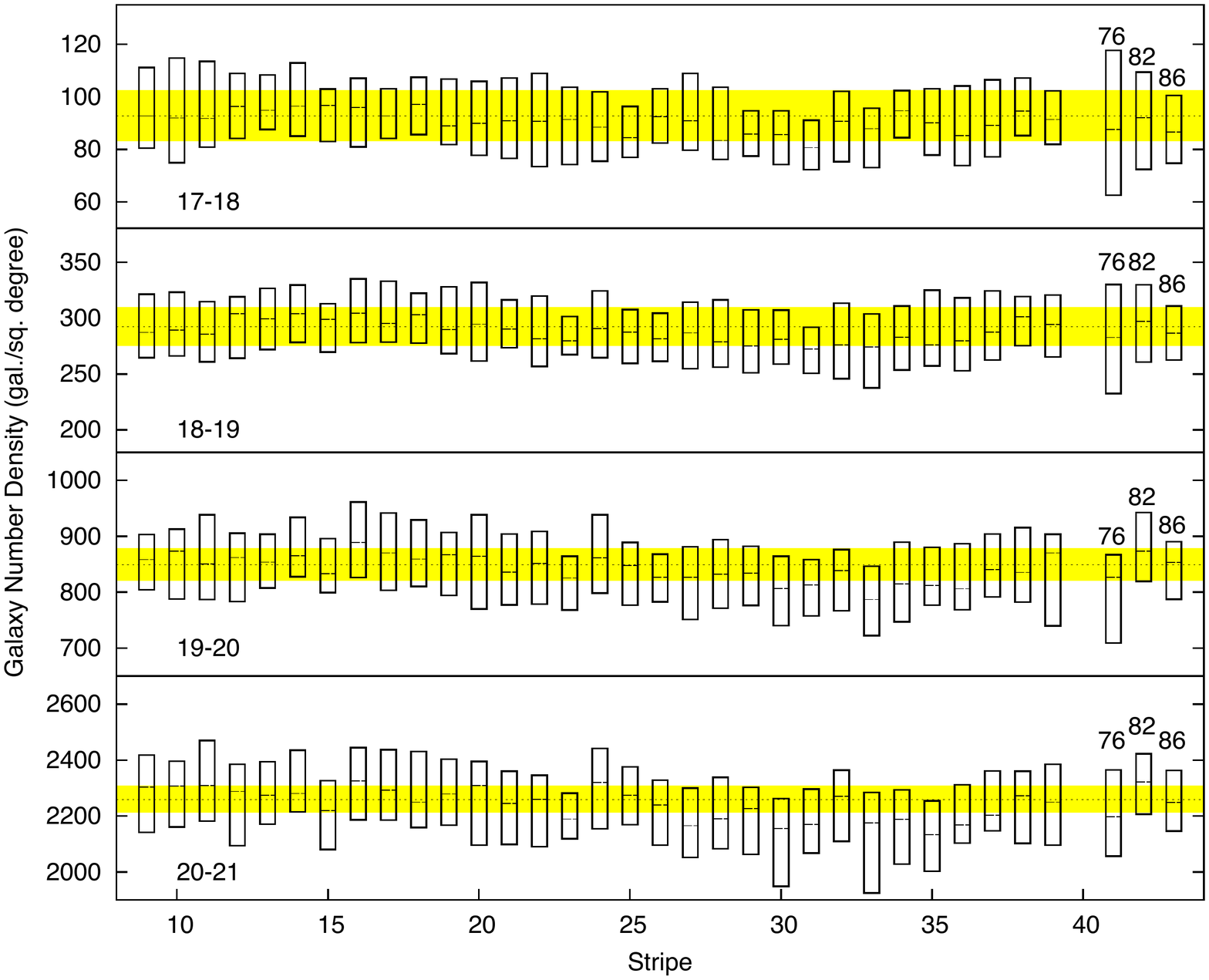}}
\end{center}
\caption{\B{Left: A box plot of the galaxy number density for each SDSS DR7 stripe (enumerated along the horizontal axis) restricted to areas of both good seeing and minimal reddening values, as defined in \S\ref{redsee}, showing the median and the $25^{\rmn{th}}$ and $75^{\rmn{th}}$ percent quartiles. The dotted line shows the mean galaxy density derived from the full main sample, which is $3493.4$ galaxies/square degree. The light yellow region shows the one sigma Poissonian variation. Right: The same box plot now divided into four magnitude ranges: $17 < r \leq 18$, $18 < r  \leq 19$, $19 < r \leq 20$, and $20 < r \leq 21$, along with their respective mean galaxy densities (shown as the dotted line) as derived from the full main sample, which are  $92.8$, $292.3$, $849.2$, and $2259.0$ galaxies per square degree, respectively.}}
\label{dataDen}
\end{figure*}

\B{Since all of these stripe observations were deemed to be photometric, we expect that star-galaxy classification (see, \eg $\S$\ref{SG}) should be consistent across all stripes. To verify this assumption,  we measured the galaxy density for each of the thirty-four stripes we use in subsequent analyses (\ie the northern stripes $9$--$39$, and southern stripes $76$, $82$, $86$). The mean galaxy density we find for the total galaxy density is 3324.0 galaxies per square degree with large density fluctuations within each stripe, while the variation between the different stripes are also significant. Similar patterns are found for the four magnitude subsamples: $17 < r \leq 18$, $18 < r  \leq 19$, $19 < r \leq 20$, and $20 < r \leq 21$, with galaxy density 88.9, 278.9, 808.9, 2147.4 galaxies per square degree respectively. One concern for these significant fluctuations would be that some fraction of these stripes have systematic effects. To test this hypothesis, we repeat this analysis by using the main galaxy sample further restricted to areas of both good seeing and minimal Galactic extinction as derived in Section~\ref{redsee}.}

\B{We present our results in Figure~\ref{dataDen}, a box plot of the galaxy density for each of these thirty-four stripes. In this type of plot, the upper and lower edges of the box indicate the $75$\% and $25$\% quartiles of the distribution and the central line indicates the median value. In this figure, the left-hand panel shows the total galaxy density for a given stripe which has been restricted to areas of both good seeing and minimal Galactic extinction. Overplotted as a dotted line is the mean galaxy density across the entire main sample, along with the one-sigma range (assuming Poissonian fluctuations), which is shown by the yellow bar. The right-hand panel presents, in a similar manner, the galaxy number density as a function of SDSS stripe for four magnitude subsamples: $17 < r \leq 18$, $18 < r  \leq 19$, $19 < r \leq 20$, and $20 < r \leq 21$.}

\B{These results indicate that the corrections made by our seeing and reddening cuts are significant. They produce number density distributions that show less variation between stripes and smaller fluctuations within each stripe, and the number densities are higher than the unmasked data. The small variations between the different stripes reflects the variation in the clustering pattern of galaxies across the sky (note that we explicitly present the clustering difference between stripes in Figure~\ref{stripesfig} in $\S$\ref{stripes}). In addition, these variations are generally consistent with random fluctuations, both in the individual magnitude ranges and the full main sample. As a result, these two systematic effects, seeing and Galactic extinction, do induce systematic signals that can be removed from our galaxy sample by using the appropriate restrictions. Since these restrictions remove the vast majority of the data from stripes $42$, $43$, and $44$, in the end we simply remove these stripes entirely from the clustering analyses of the main galaxy sample.}

\subsection{\em Seeing Variations}\label{see}

\begin{figure}
\begin{center}
\resizebox{8cm}{!}{\includegraphics{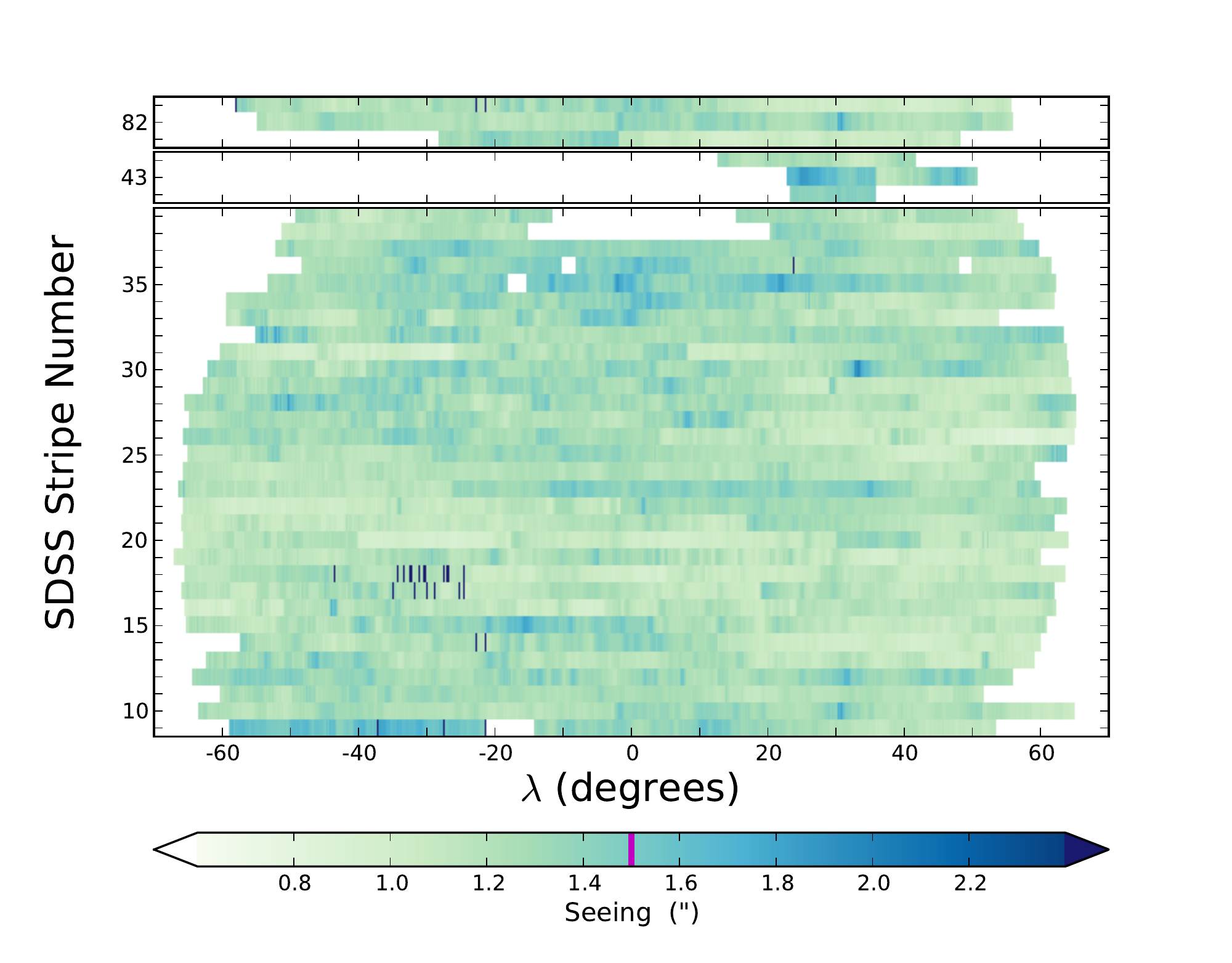}}
\end{center}
\caption{{A heat map showing the average seeing values as a function of the SDSS lambda (\ie longitude) coordinate for all thirty-seven stripes in the SDSS DR7. The bottom panel shows the northern hemisphere stripes $9$--$39$, while the top panel shows stripes $42$--$44$, and the southern hemisphere stripes: $76$, $82$, and $86$. For convenience, the three southern hemisphere stripes are shifted in lambda to align with northern stripes. The final value we use to remove the systematics from seeing is indicated in the colorbar at the bottom of the figure with a vertical magenta line.}}
\label{seeVsLam}
\end{figure}

To determine the seeing as a function of spatial location, we use the effective area of the point-spread function for each measured survey field to determine the relevant seeing values\footnote{\url{http://www.sdss.org/dr7/algorithms/masks.html}}. As described in $\S$\ref{CrossFunMethod}, we pixelate the entire SDSS DR7 footprint by using SDSSPix at resolution 128, and assign each pixel the appropriate stripe number, the $\lambda/\eta$ coordinate of the pixel center, and the relevant seeing and reddening values. We present the calculated seeing values as a function of lambda (\ie the SDSS longitude coordinate) for each stripe in the SDSS DR7 northern contiguous region and the three separate southern stripes in Figure~\ref{seeVsLam}. The bottom panel contains the contiguous northern hemisphere stripes $9$--$39$, while the top panel contains the northern stripes $42$--$44$ and the three southern stripes: $76$, $82$, and $86$. Overall, the seeing for all stripes generally remains fairly smooth, as expected, with most seeing values below $1\farcs5$. By using this as a canonical value, only stripe $43$ was observed primarily in less than ideal conditions.

\begin{figure}
\resizebox{15 cm}{!}{\includegraphics{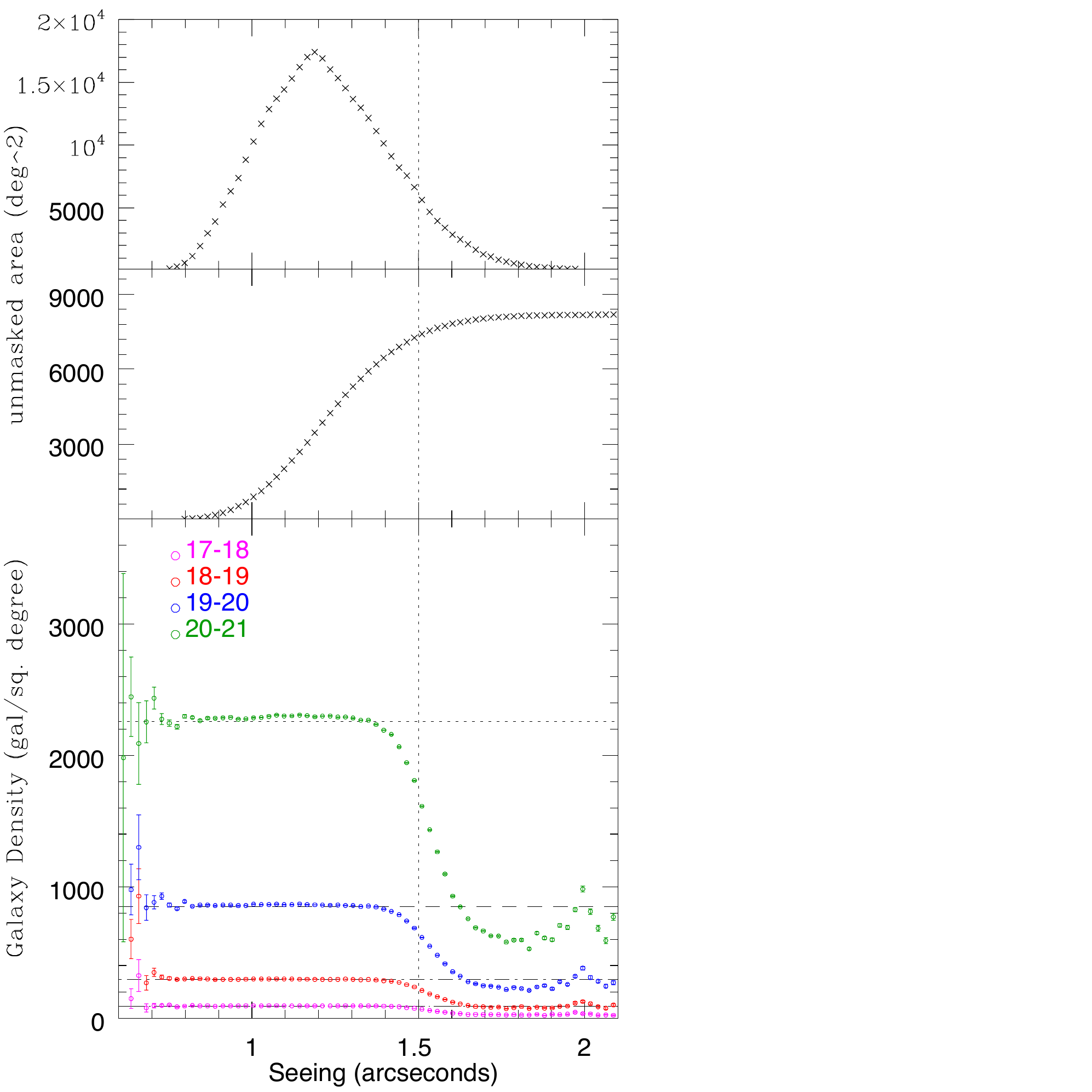}}
\caption{\AC{Top: The differential unmasked area as a function of seeing. Middle: The cumulative unmasked area to the total survey area as a function of seeing. Bottom: The differential galaxy number density as a function of seeing. The four horizontal lines are the mean densities of the full sky coverage for four magnitude bins from the right panel of Figure~\ref{dataDen}. The error bar are Poissonian fluctuations in each seeing bin. The vertical dot line shows for the seeing cut that we use for our final galaxy catalog.}}
\label{seeAreaGalDen}
\end{figure}

In general, we want to both minimize the effect of a systematic on our clustering measurements while maximizing the number of sources (or equivalently observed area) available for analyses. Using the pixelised SDSS DR7 map, we calculate the survey area as a function of seeing, which we display as the differential area in the top panel of Figure~\ref{seeAreaGalDen}, and as the cumulative area in the middle panel of Figure~\ref{seeAreaGalDen}. From this figure, we see that pixels with seeing values smaller than $1\farcs2$ contain approximately half of the total observed area, while the pixels with seeing values smaller than $1\farcs5$ contain almost the entire observed area. As a result, the majority of the survey area will be retained by using a seeing cut between $1\farcs2$ and $1\farcs5$.

\B{Next, we explore how the galaxy number density depends on the seeing. To obtain these values, we augment our pixelised SDSS DR7 map with the galaxy density for each pixel. We plot the binned, differential galaxy number counts as a function of seeing in four magnitude ranges in the bottom panel of Figure~\ref{seeAreaGalDen}. In this figure, the galaxy densities have large fluctuations at small seeing values while this fluctuation quickly decreases as we include more area. By looking at this figure in conjunction with the differential area figure in the top panel, we can see that the galaxy density at low seeing values oscillates due to the small number of pixels with very small seeing values. Likewise, we see that the increase in the variation of the galaxy number density at higher seeing values occurs since there are few pixels with higher seeing values.}

\B{As shown in the figure, the galaxy number density decreases at large seeing values. At seeing value of $\sim1\farcs5$, the differential galaxy number density is 80\% of the density at smaller seeing. This decrease can be understood since as the seeing increases, star/galaxy classification becomes more difficult due to the atmospheric blurring of the source light profiles. This effect decreases the galaxy number counts in each pixel; and, therefore, decreases the overall galaxy density. By adopting differential galaxy number densities higher than 80\%, this figure indicates that a maximum seeing cut at $1\farcs5$ should be used; however, the exact value to be used is best determined by a cross-correlation measurement as discussed in Section~\ref{redsee}.}

\subsection{\em Reddening Variations}\label{red}

\begin{figure}
\begin{center}
\resizebox{8cm}{!}{\includegraphics{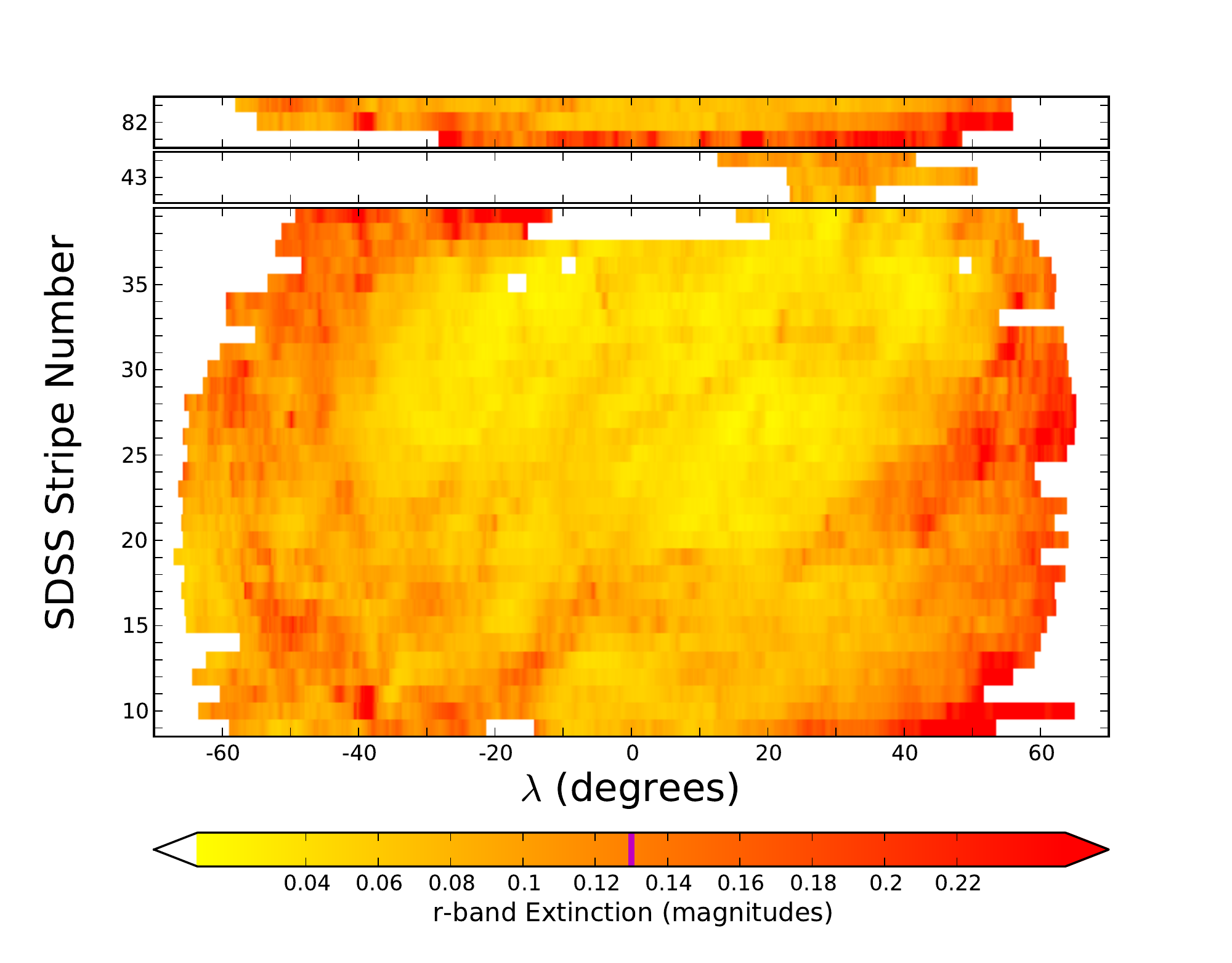}}
\end{center}
\caption{{A heat map showing the average reddening values as a function of the SDSS lambda (\ie longitude) coordinate for all thirty-seven stripes in the SDSS DR7. The bottom panel shows the northern hemisphere stripes $9$--$39$, while the top panel shows stripes $42$--$44$, and the southern hemisphere stripes: $76$, $82$, and $86$. For convenience, the three southern hemisphere stripes are shifted in lambda to align with northern stripes. The final value we use to remove the systematics from reddening is indicated in the colorbar at the bottom of the figure with a vertical magenta line.}}
\label{redVsLam}
\end{figure}

Galactic extinction (or reddening) systematically dims objects, and the spatial distribution of the dust that causes this obscuration within our Galaxy varies across the sky. Thus, to determine an acceptable limit for this systematic effect, we follow a similar procedure to the one outlined in the section~\ref{see} where we pixelate the sky (as described in Section~\ref{CrossFunMethod}). In this case, however, we start by using the reddening map of~\citet{SFD} to quantify the Galactic extinction as a function of the SDSS lambda coordinate for each stripe in the SDSS DR7 northern hemisphere and the three separate southern stripes, as shown in Figure~\ref{redVsLam}. These two observed regions are centered near the northern and southern Galactic poles, which are both regions of low Galactic extinction. We, therefore, expect \textit{a priori} that all of these stripes should generally have higher reddening values at their endpoints in comparison to their midsection, which is the trend that is generally seen in Figure~\ref{redVsLam}.

\begin{figure}
\begin{center}
\resizebox{15 cm}{!}{\includegraphics{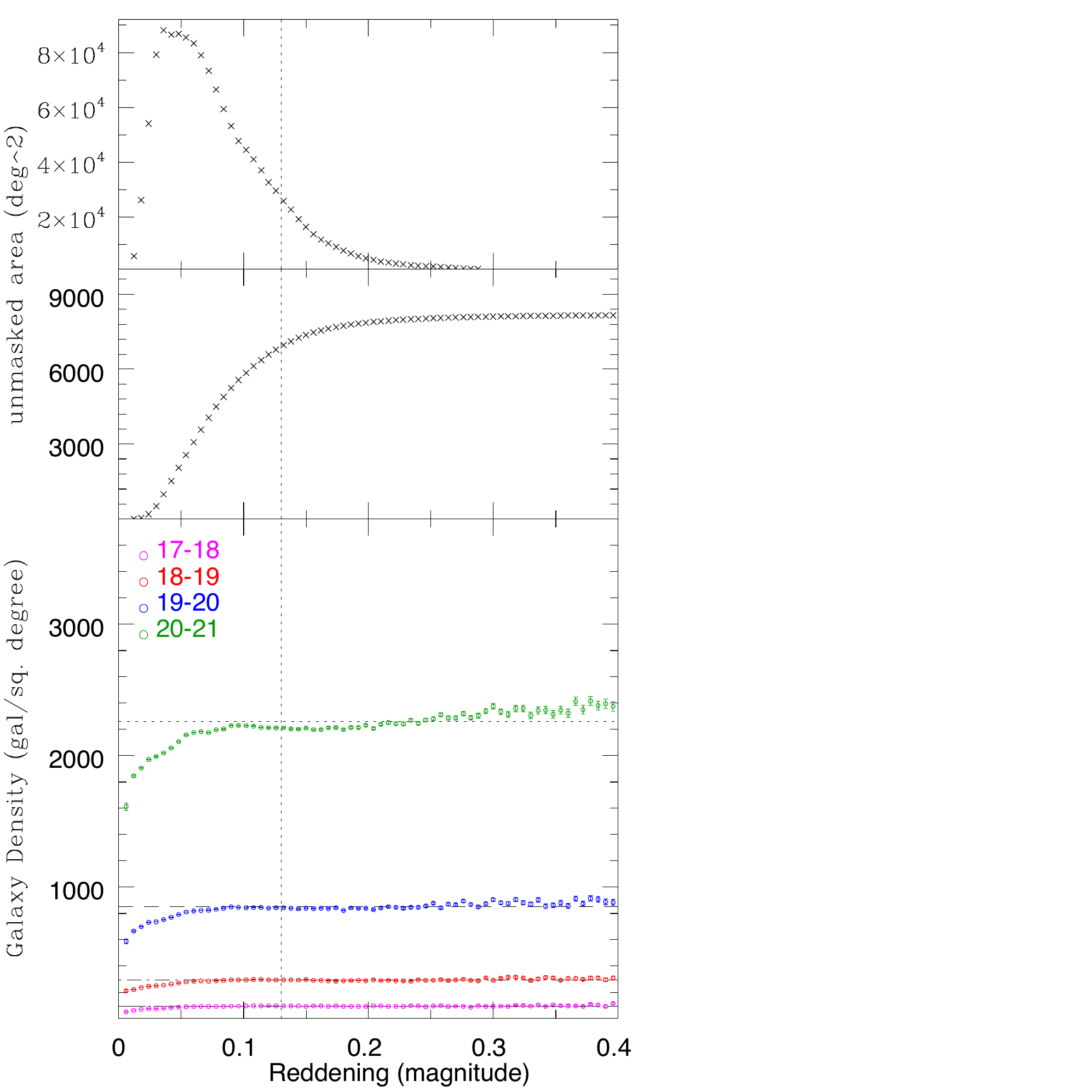}}
\end{center}
\caption{\AC{Top: The differential unmasked area as a function of reddening. Middle: The cumulative unmasked area to the total survey area as a function of reddening. Bottom: The differential galaxy number density as a function of reddening. Similar as Figure~\ref{seeAreaGalDen}, the four horizontal lines are the mean densities of the full sky coverage for four magnitude bins from the right panel of Figure~\ref{dataDen}. The error bar are Poissonian fluctuations in each seeing bin. The vertical dot line shows for the reddening cut that we use for our final galaxy catalog.}}
\label{redAreaGalDen}
\end{figure}

As discussed in $\S$\ref{see}, we want to maximize the retained survey area, while minimizing the effects of the systematic, in this case Galactic extinction, on our clustering measurements. Using this pixelised reddening map, we calculate the survey area as a function of reddening, which we display as the differential area in the top panel of Figure~\ref{redAreaGalDen}, and as the cumulative area in the middle panel of Figure~\ref{redAreaGalDen}. From this figure, we see that pixels with reddening values  less than $0.1$ include nearly $75$\% of the survey area, while reddening values less than $0.2$ include nearly all of the survey. As a result, the majority of the survey area can be maintained by using a reddening cut between $0.1$ and $0.2$.

\B{Next, we explore how the galaxy number density varies with Galactic extinction.  We plot the binned galaxy number density as a function of reddening in four magnitude ranges in the bottom panel of Figure~\ref{redAreaGalDen}. For all magnitude ranges, the scatter in the distribution increases for reddening values larger than $0.2$, indicating that there are few pixels with reddening values in this range. On the other hand, at small reddening values, \ie below $0.1$ magnitudes, the galaxy density increases as the reddening value increases. As the value increases, the amount of survey area included also increases, and we eventually reach a nearly steady galaxy density around a reddening value of $0.1$. We therefore conclude that we will want to make a reddening cut somewhere between $0.1$ and $0.2$, but once again we will quantify the exact value by using a cross-correlation measurement as discussed in Section~\ref{redsee}.} 

\subsection{\em Cross-Correlations - Galaxy Density against Seeing and Reddening}\label{redsee}

In the previous two subsections, we determined the optimal ranges for the values of both seeing and Galactic extinction that would minimize their systematic effects on our correlation measurements. In this section, we now focus on determining the actual values for each of these systematic effects, which we accomplish by calculating the galaxy-seeing and galaxy-reddening cross correlation functions. To measure these correlation functions, we first pixelate the sky so we can calculate the pixel cross-correlation function as described in Section~\ref{CrossFunMethod}. Ideally, we can identify a systematic value that produces a flat cross correlation function that is consistent with zero on both small and large scales. In practice, some residual will remain; therefore we measure the cross-correlation function for different values of each systematic in order to find the optimal value.

\subsubsection{\em Cross-Correlation Function Estimators}\label{CrossFunMethod}

\begin{figure*}
\begin{center}
\resizebox{8cm}{!}{\includegraphics{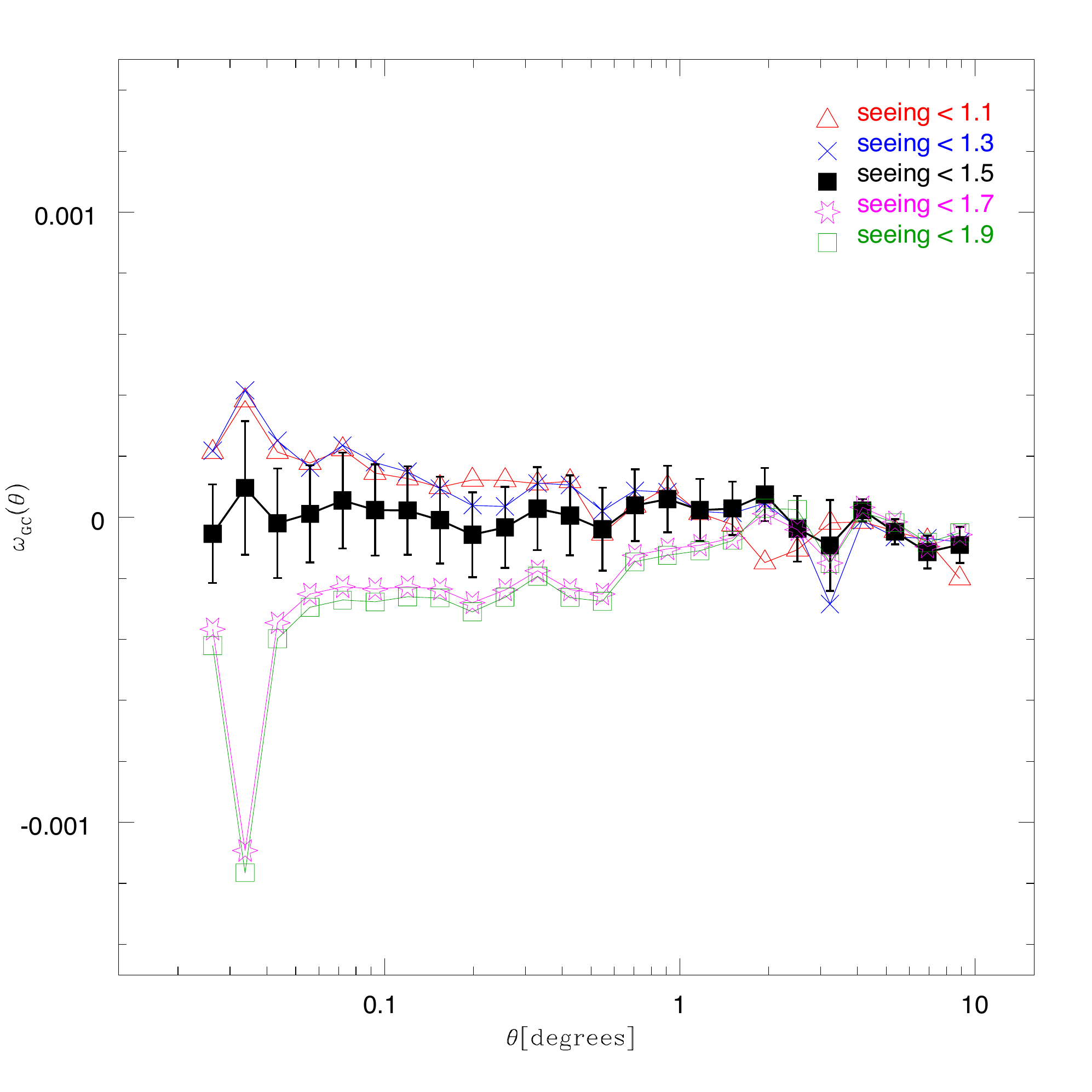}}
\resizebox{8cm}{!}{\includegraphics{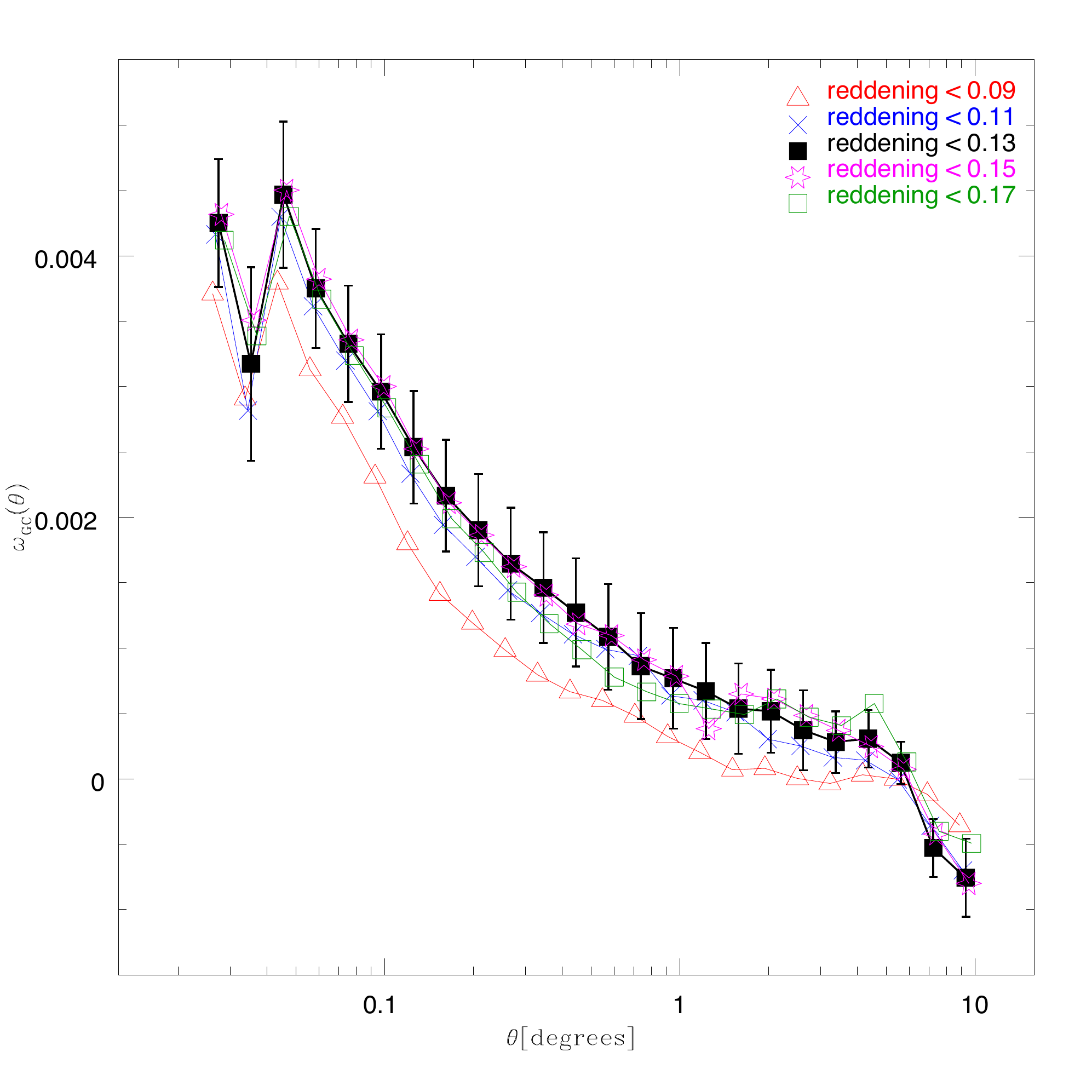}}
\end{center}
\caption{\B{Left: The galaxy-seeing cross-correlation functions for $17 < r \le 21$. The bolded black square points and error bars represent the preferred seeing cut of $1\farcs5$. Right: The galaxy-reddening cross-correlation functions for $17 < r \le 21$. The bolded black square points and error bars represent the preferred reddening cut of $0.13$. The error bars in two panels are typical for the correlation functions calculated by using the other seeing or reddening values. }}
\label{cross17to21}
\end{figure*}

To determine the optimal data sample for our analysis, we need to quantify the specific data cuts we employ to minimize systematic effects on our measurement. In particular, we wish to minimize the effects of seeing and Galactic extinction, or reddening. As demonstrated by~\citet{Scr}, this can be accomplished by measuring the two-point angular cross-correlation function between galaxies and the relevant systematic. As both reddening and seeing are not observed as continuous quantities, however, we must first pixelate the sky by using the HEALPix scheme as described in Section~\ref{pixelisation}. The main caveat with this approach is that to measure cross-correlations for these systematics, we must adopt pixels that are smaller than the characteristic scale of the observed systematic effect. Because each SDSS scan line has approximately $0\fdg21$ in width and $160\degr$ in length, we expect the systematic effect due to seeing to be bounded by the width of a single SDSS scan line, which should also be less than the image frame size (the frame size is described at \url{http://www.sdss.org/dr7/instruments/imager/}, and is about $0.0337$ square degrees). The reddening values published by the SDSS are derived from the~\citet{SFD} maps, which have an even larger pixel size.
Thus, the minimum pixel area we use for our cross-correlation measurements must be less than the image frame size, or $0.0337$ sq. deg. As a result, we use HEALPix resolution 2048 to pixelate the SDSS DR7 data, which corresponds to a pixel size of 0.00082 square degrees.

We next compute both the number of galaxies and the mean seeing and reddening values for each pixel. Following~\citet{Scr}, we divide the entire SDSS DR7 data into $10^{\circ}\times10^{\circ}$ subsamples, and measure the mean number density of galaxies per pixel and the mean systematic per pixel for each of these subsamples. Using the galaxy counts and mean systematic values, we calculate the over/under density for both the number of galaxies and the systematic for each pixel $i$ within a specific subsample:
\begin{equation}
\begin{aligned}
\delta^g_i &= \frac{n^g_i-\bar{n}^g}{\bar{n}^g}, \\
\delta^s_i &= \frac{v^s_i-\bar{v}^s}{\bar{v}^s},
\end{aligned}
\end{equation}
where $n^g_i$ is the galaxy number density (indicated by $g$) for pixel $i$, and $v^s_i$ is the mean value of the systematic being quantified (\eg seeing or reddening, indicated by $s$) for pixel $i$. $\bar{n}^g$ and $\bar{v}^s$ are the mean galaxy number density per pixel and the mean value of the specific systematic for the given subsample, respectively.

By using these pixelised quantities, we use the following estimator to calculate the angular cross-correlation of galaxies against a specific systematic quantity:
\begin{equation}
\omega(\theta)=\frac{\sum_{i,j} \delta^g_i\delta^s_j\Theta_{ij}}{\sum_{i^*,j^*} \Theta_{i^*j^*}}.
\end{equation}
If the distance between $i$ and $j$ are within the given $\theta$ bin, $\Theta_{ij}$ is equal to one, otherwise it is zero. The estimator is calculated between 0$^\circ$.05 and 10$^\circ$, with a logarithmic scale of 30 angular bins. Once the estimator has been calculated for all subsamples, we calculate the mean estimator $<\bar{\omega}(\theta)>$ and the error from all subsamples by using the following equation:
\begin{equation}
(\delta\omega(\theta))^2=\frac{1}{N^2}\sum_{n=1}^N(\bar{\omega}(\theta)-{\omega}_i(\theta))^2,
\end{equation}
where $N \sim 100$, which is how many subsamples we use in this measurement.

\subsubsection{\em Results}\label{CrossResults}

\begin{figure}
\begin{center}
\resizebox{8 cm}{!}{\includegraphics{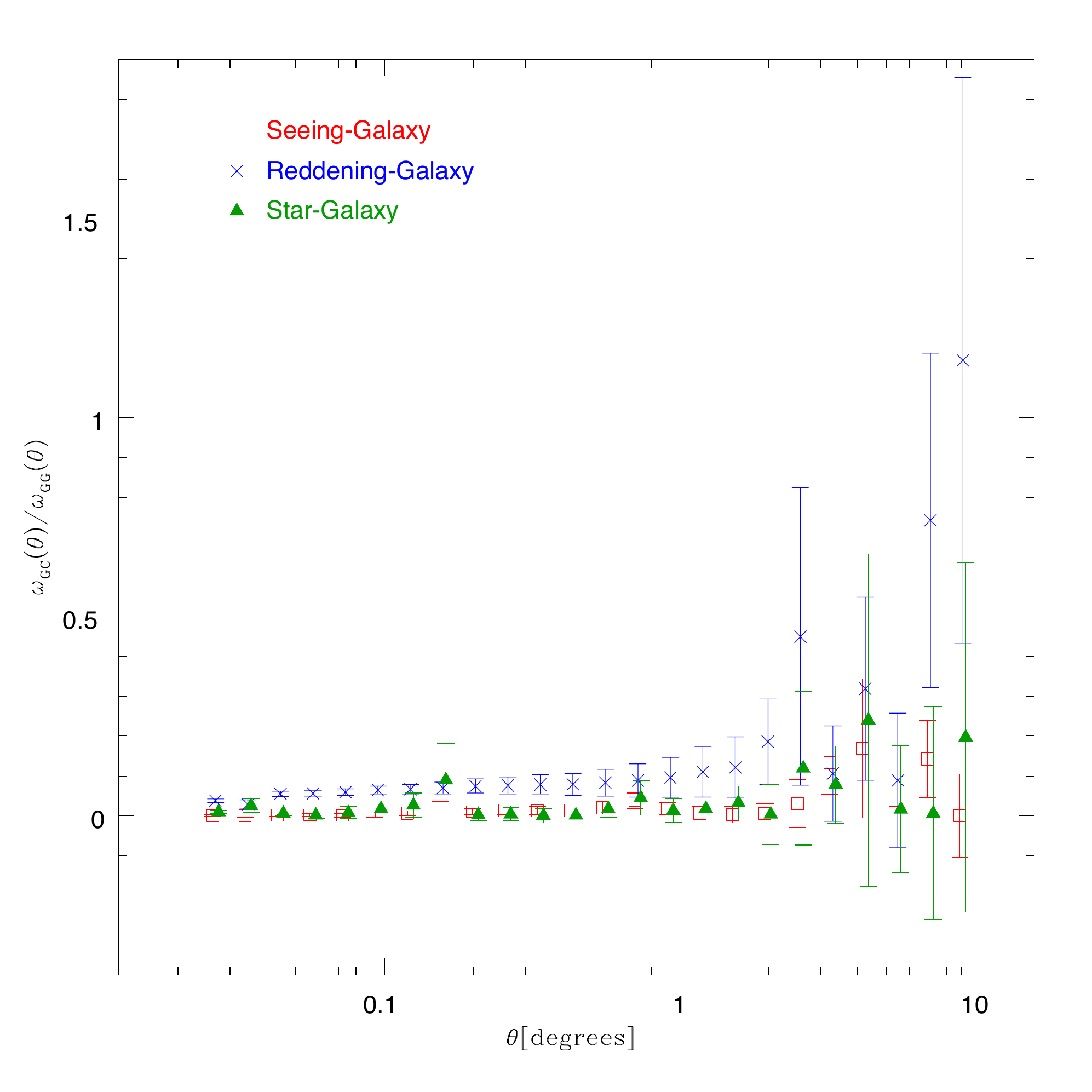}}
\end{center}
\caption{\B{The galaxy-galaxy auto-correlation function compared to the galaxy-reddening, galaxy-seeing, and galaxy-star cross-correlation functions for galaxies and stars with magnitudes in the range $17 < r \le 21$ with seeing $< 1\farcs5$ and reddening $< 0.13$. These systematic signals are well below the galaxy auto-correlation function until $\sim5\degr$.}}
\label{galcross}
\end{figure}

\B{In the left panel of Figure~\ref{cross17to21}, we present the galaxy-seeing cross-correlation function for the full sample over the magnitude range $17 < r \le 21$. We calculated the pixel cross-correlation function for seeing values between $1\farcs0$ and $2\farcs0$ in steps of $0\farcs1$, but only show the five correlation functions for clarity (the other samples show similar trends). This figure indicates that seeing cuts at or smaller than $1\farcs5$ have minimal systematic effects as the cross-correlation function is mostly consistent with zero, especially at large scales. Since a seeing cut of $1\farcs5$ keeps more than $90$\% of the survey data while minimizing the contamination cross-correlation signal, we choose $1\farcs5$ as the final value of our seeing cut. Figure~\ref{galcross} indicates that this signal is much less than the galaxy auto-correlation function measurement $\omega(\theta)$ on all scales, from $0\fdg05$ to $\sim5\degr$.}

\B{Likewise, in the right panel of Figure~\ref{cross17to21}, we present the galaxy-reddening cross-correlation function for the full sample over the magnitude range $17 < r \le 21$. The reddening cross-correlation function is calculated for both magnitude samples from $0.1$ to $0.2$ magnitudes in intervals of $0.01$ magnitudes. However, for clarity only the five correlation functions are shown (again, the others follow similar trends). The reddening cuts are all consistent with zero within 3$\sigma$ at small scales and within 1$\sigma$ at large scales. We are especially interested in the large angle cross-correlation function values ($\sim 5\degr$, where the reddening cross-correlation signal is of similar scale to the galaxy correlation). Therefore, we choose the reddening cut that has the smallest value at $5\degr$, while also keeping the majority of the survey area. As a result, we select $0.13$ magnitudes to be the upper limit for our allowed reddening value, which keeps more than $80$\% of the data. We note that as shown in Figure~\ref{galcross}, the galaxy-reddening cross-correlation signal is well below the value of the galaxy auto-correlation function until around $5\degr$.}

\B{We also measure the cross-correlation functions for both galaxy-seeing and galaxy-reddening in the four magnitude bins, and find similar trends with the full sample. We also measure the galaxy-star cross-correlation function, which is below the galaxy auto-correlation function until $\sim 5\degr$. In Figure~\ref{galcross}, we show the ratio of  the pixelized galaxy-seeing, galaxy-reddening, and galaxy-star cross-correlation functions to the pixelized galaxy autocorrelation function. We find $\sim 5\degr$ is the scale where both the reddening and star galaxy cross-correlation functions become comparable in magnitude with the galaxy auto-correlation function, while the galaxy-seeing cross-correlation is always well below the galaxy signal with small error bars. We discuss the galaxy-star cross-correlation function in more detail in $\S$\ref{Conclusion}.}

\begin{figure}
\begin{center}
\resizebox{8 cm}{!}{\includegraphics{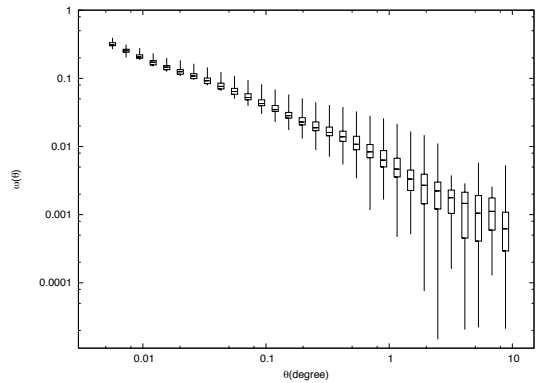}}
\end{center}
\caption{\B{A box-whisker plot of the stripe galaxy angular correlation functions in the magnitude range $17 < r \le 21$ for the thirty-one northern stripes $9$--$39$, and three southern stripes: $76$, $82$, and $86$.}}
\label{stripesfig}
\end{figure}

\subsection{\em Correlation Function Among Stripes}\label{stripes}

Having applied the systematic cuts for reddening and seeing, we now complement the technique discussed in Section~\ref{numDenMethod} to verify the uniformity of our final galaxy sample across SDSS stripes. We measure the galaxy auto-correlation functions by using the point-to-point technique discussed in $\S$\ref{Code} for each individual stripe to quantify the stripe-to-stripe fluctuations. In Figure~\ref{stripesfig}, we present a box-whisker plot for the galaxy auto-correlation functions of the thirty-one northern stripes $9$--$39$, and three southern stripes: $76$, $82$, and $86$. In this type of plot, the box shows the span of the central $50$\% of the data while the whiskers show the minimum and maximum limits of the data (in this case the galaxy auto-correlation function across all stripes at a given angular resolution). 

As indicated by the whiskers, there are some variations across the different stripes, which is expected since the clustering of galaxies will vary across the sky. We see exactly this type of variation in the density of galaxies across the same SDSS stripes as shown in Figure~\ref{dataDen}. Taken together, these results provide evidence that our final, masked galaxy sample is sufficiently uniform across the specified SDSS footprint for our angular correlation analysis.

\section{The SDSS Galaxy Angular Correlation Function}\label{Correlation}

\subsection{\em The Angular Correlation Function Estimator}

After concluding the systematic tests and defining the final galaxy sample as detailed in Section~\ref{SysResults}, we next focus on measuring the clustering of the galaxy sample by using the two-point galaxy angular correlation function. The two-point galaxy angular correlation function calculates the excess probability over a random distribution that given one galaxy at a specific location, another galaxy will be found within a specific angular distance~\citep{Peebles}. Given such a probabilistic definition, it is not surprising that to determine this function we require a large number of random points. Therefore, we construct a large random sample of galaxies (the total number of random points used in any measurement is always at least ten times the size of the individual galaxy sample being analysed) that both lie within the SDSS theoretical footprint and that are also restricted to areas of the sky that satisfy the systematic cuts discussed in the Section~\ref{redsee}.

With these random points, we measure the two-point galaxy correlation function by using the~\citet{Land} estimator:
\begin{equation}
\omega(\theta)=\frac{N_{dd}-2N_{dr}+N_{rr}}{N_{rr}},
\label{landEq}
\end{equation}
where $N_{dd}$ is the normalized number of galaxy-galaxy pairs counted within a given angular separation bin of $\theta \pm \delta\theta$ (\eg over the entire SDSS DR7 galaxy sample), and $N_{dr}$ and $N_{rr}$ are the normalized number of galaxy-random pairs and random-random pairs, respectively. Unless stated otherwise, we calculate the two-point galaxy angular correlation function in thirty angular bins, spaced logarithmically between $0\fdg005$ and $10\degr$.

\subsubsection{\em Different Correlation Estimators}

\begin{figure}
\begin{center}
\resizebox{8 cm}{!}{\includegraphics{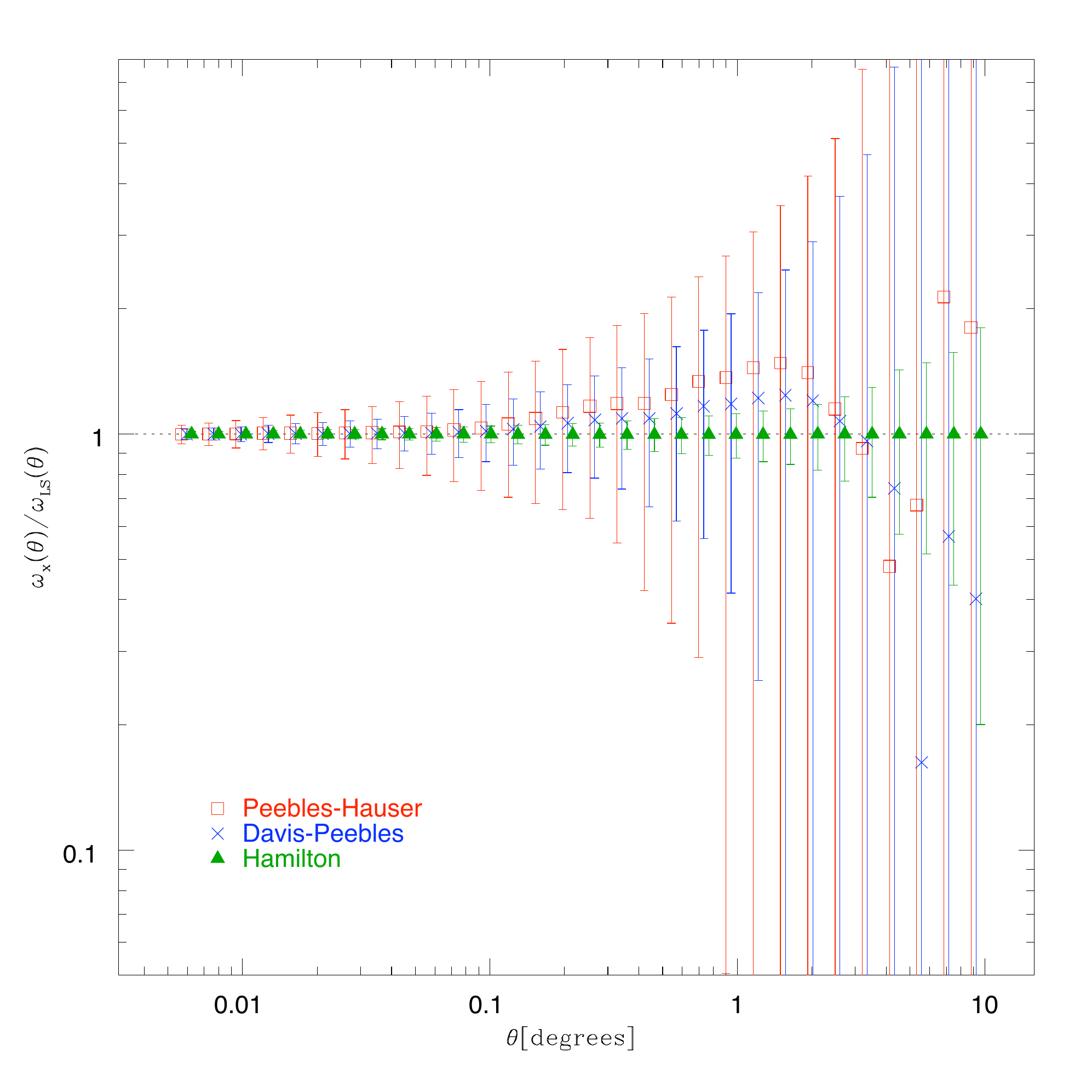}}
\end{center}
\caption{{The ratio of the~\citet{Peebles74} estimator (open squares), the~\citet{Davis} estimator (crosses), and the~\citet{Hamil} estimator (filled triangles) to the~\citet{Land} estimator. The errors for each estimator are calculated by using 32 jackknife resamplings.}}
\label{estiComFig}
\end{figure}

Besides the~\citet{Land} estimator presented in the previous section, we have explored three other estimators for the two-point angular correlation function. First, we have tried the original Peebles estimator~\citep{Peebles74}:
\begin{equation}\label{peeblesEq}
\omega(\theta)_{PH}=\frac{DD/N^2_D-RR/N^2_R}{RR/N^2_R}.
\end{equation}
Second, we have tried a similar estimator developed by~\citet{Davis}:
\begin{equation}\label{DavisEq}
\omega(\theta)_{DP}=\frac{DD/N_D-DR/N_R}{DR/N_R}.
\end{equation}
Finally, we have tried the following estimator developed by~\citet{Hamil}:
\begin{equation}\label{hamiltonEq}
\omega(\theta)_{H}=\frac{DD*RR-DR*DR}{DR*DR}.
\end{equation}
In all three of the above equations, DD is the galaxy-galaxy pair counts within the given angular bin, while DR and RR represent the bin counts of galaxy-random pair and random-random pair, respectively. Likewise, N$_D$ and N$_R$ are the total number points in the galaxy sample and random sample and are used to properly normalize the appropriate pair count.

We compare these three estimators with the standard~\citet{Land} estimator in Figure~\ref{estiComFig}. As has been shown previously~\citep{Ker00}, the~\citet{Hamil} estimator is in close agreement with the~\citet{Land} estimator, but has slightly larger error bars. On the other hand, both the~\citet{Peebles74} and the~\citet{Davis}  overestimate the galaxy clustering at small scales and have larger error bars over all scales than the~\citet{Land} estimator. As a result, we will utilize the~\citet{Land} estimator as appropriate throughout this paper.

\subsection{\em Errors and Curve Fitting}
\begin{figure*}
\begin{center}
{\resizebox{8 cm}{!}{\includegraphics{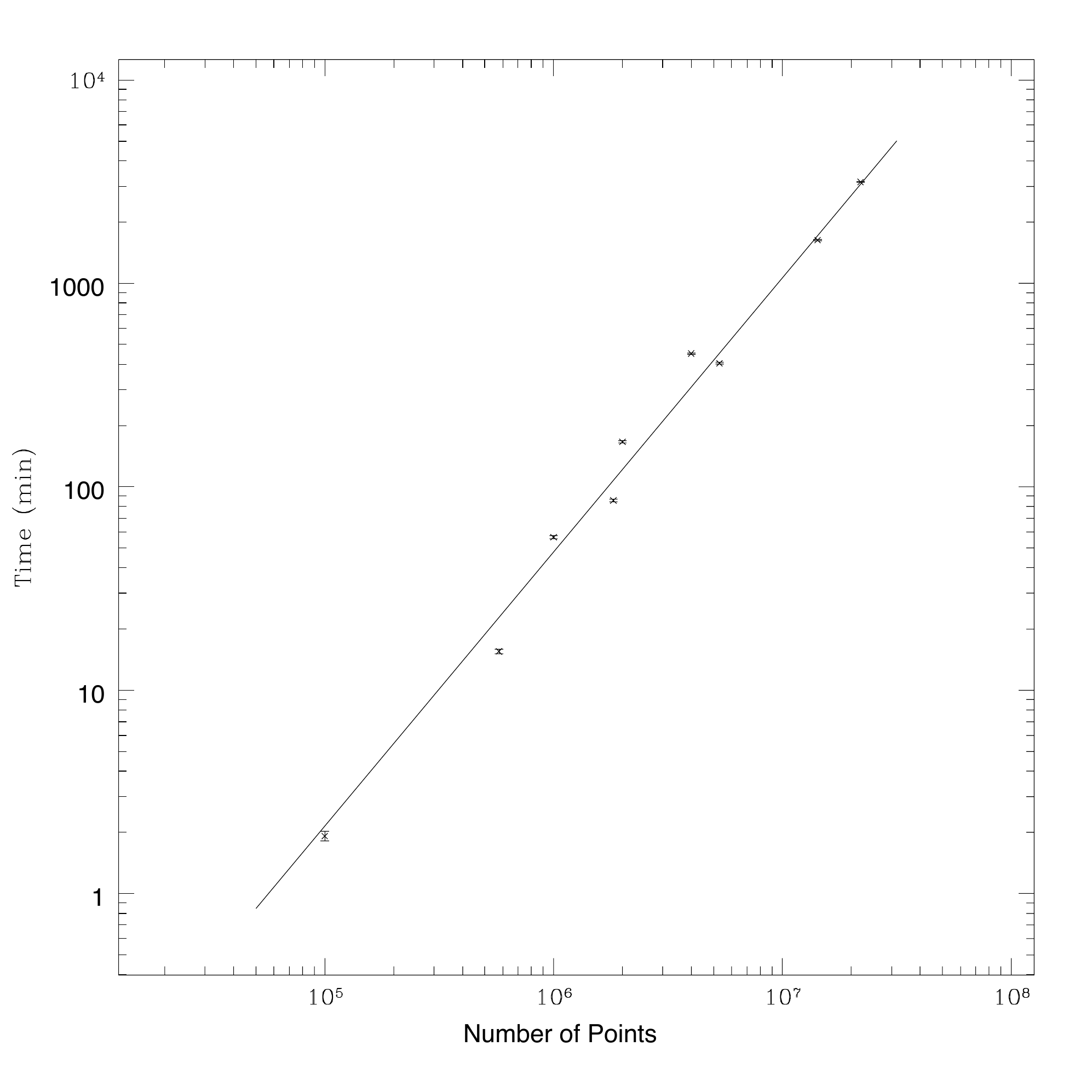}}}
{\resizebox{8 cm}{!}{\includegraphics{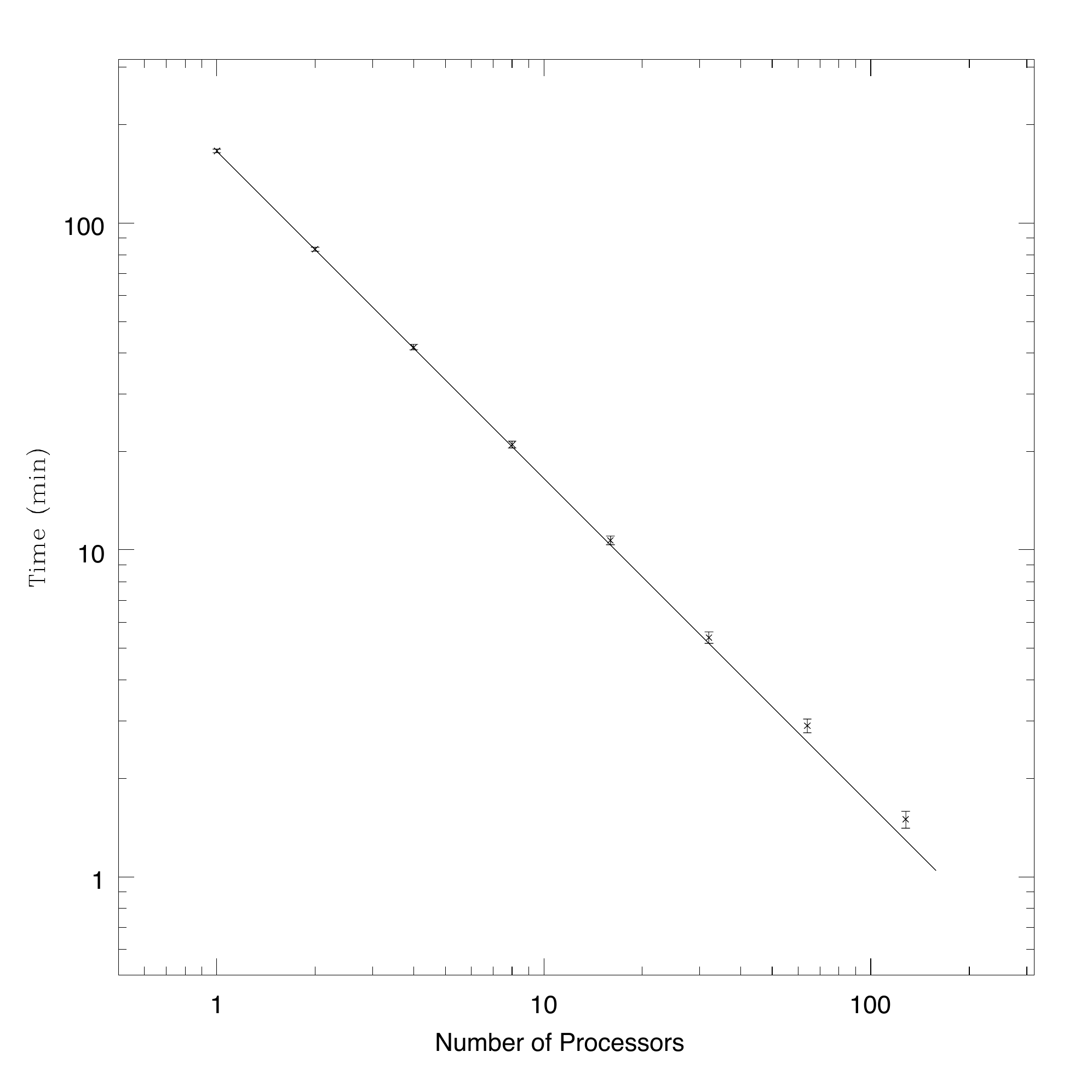}}}
\end{center}
\caption{{Left: Computational time for the calculation of the two-point galaxy auto-correlation function by using our specialized code as a function of the number of galaxies. The fitted line shows the runtime scales with $N^{1.35}$. Right: The computational time as a function of the number of processors, for approximately 2 million galaxies. The line shows the ideal runtime as processors increases.}}
\label{timescale}
\end{figure*}

To calculate the errors on our two-point galaxy angular correlation function measurements, we adopt the `delete one jackknife' method. We subdivide our full galaxy sample into 32 sub-samples. By leaving one subsample out, we calculate the two-point galaxy angular correlation function for the data in the remaining thirty-one subsamples. This allows us to construct a jackknife defined covariance matrix that both quantifies the homogeneity of our galaxy sample and also allows us to optimally model-fit our correlation function measurements. 

The covariance matrix for the N = 32 jackknife samples is determined by using the formula presented by~\citet{Scr}, but see also~\citet{Zehavi02, Myers05, Myers07}:
\begin{equation}
C(\theta_i, \theta_j) = \frac{31}{32} \sum_{k=1}^{32} [\omega(\theta_i) - \omega_k(\theta_i)] [\omega(\theta_j) - \omega_k(\theta_j)],
\label{covmat}
\end{equation}
where $\omega$ is the value from the full galaxy sample, the $\omega_k(\theta)$ refers to the value of the correlation measurement obtained by omitting the $k^{\textrm{th}}$ subsample of data, and $i$ and $j$ are respectively the $i^{\textrm{th}}$ and $j^{\textrm{th}}$ angular bins. The jackknife bin errors, $\sigma_i$, can be obtained from the diagonal elements of the covariance matrix, \ie
\begin{equation}
\sigma_i^2 = C_{i,i}.
\end{equation}

For comparison with previous works~\citep[\eg][]{Con}, we fit our two-point galaxy angular correlation measurements with a power-law model: $\omega_m(\theta) = A_\omega \theta^{(1-\gamma)}$. To determine the best fit model for each correlation function, we perform a chi-squared minimization~\citep{Press}:
\begin{equation}
\chi^2 = \frac{1}{N_{dof}} \sum_{i,j} [\omega(\theta_i) - \omega_m(\theta_i)] C_{i,j}^{-1} [\omega(\theta_j) - \omega_m(\theta_j)],
\label{curfit}
\end{equation}
where $\omega(\theta)$ is the measured two-point galaxy angular correlation function, $\omega_m(\theta)$ is the model two-point galaxy angular correlation function, and $C_{i,j}$ are the elements of the calculated covariance matrix from Equation~\ref{covmat}.

\subsection{The Fast Two-Point Correlation Function Calculation}\label{Code}

Historically, the two-point correlation function (hereafter 2PCF) has been limited by the availability of large data sets with sufficient sky coverage and depth to provide a fair sample of objects in the Universe. We now live in a privileged era when such data sets are or will be available thanks to current or planned large-scale surveys such as the SDSS, the Dark Energy Survey (DES), or the Large Synoptic Survey Telescope (LSST). With millions or possibly billions of unique objects, the traditional methods of calculating the 2PCF become entirely unfeasible as calculation times quickly reach years or longer. We therefore present a technique that leverages a two-dimensional quad-tree structure to speed up these calculations. Detailed discussion of this technique can be found in~\citep{Dol}, and we make our implementation freely available\footnote{\url{http://lcdm.astro.illinois.edu/code}}.

\subsubsection{\em The Methodology}

\begin{figure*}
\begin{center}
{\resizebox{8 cm}{!}{\includegraphics{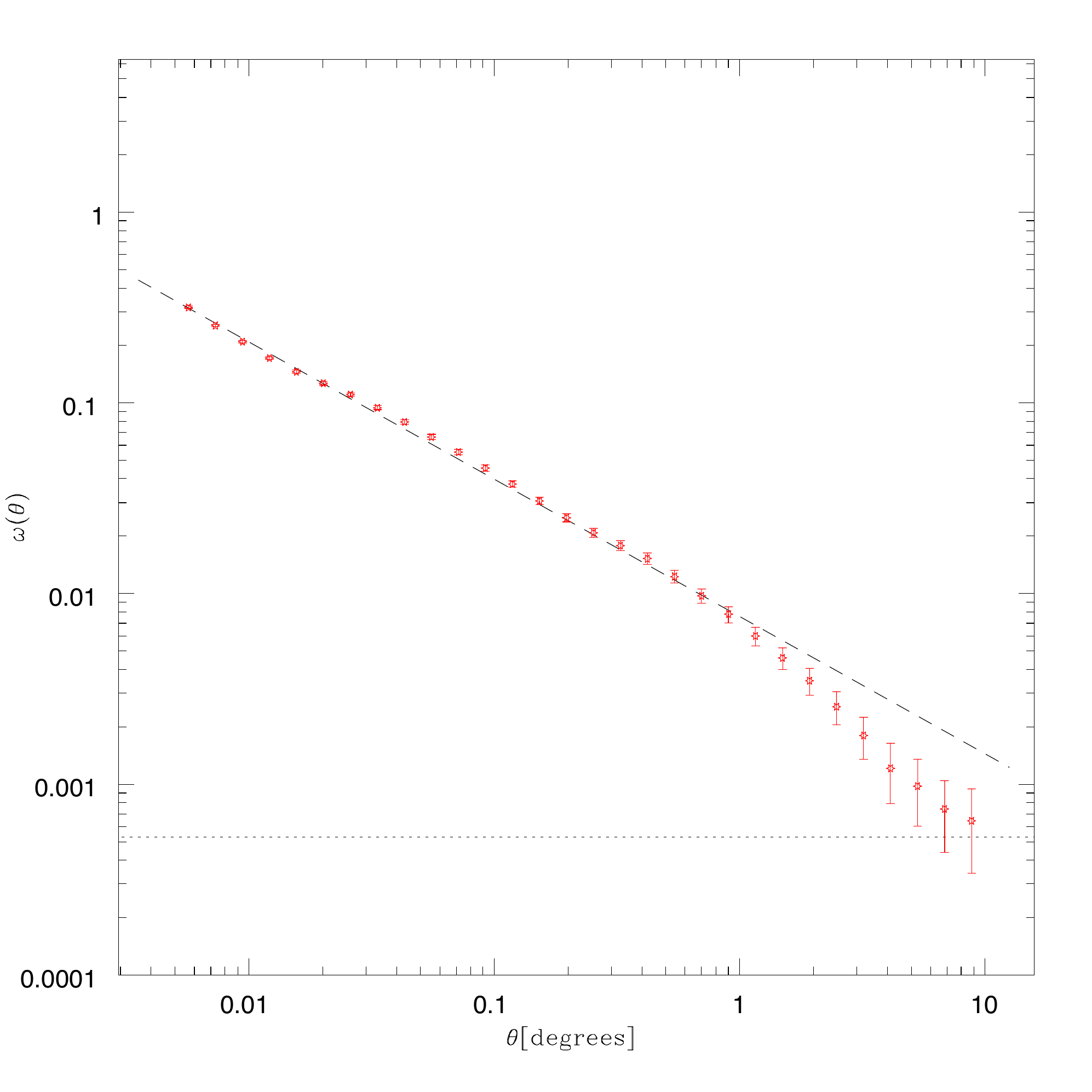}}}
{\resizebox{8 cm}{!}{\includegraphics{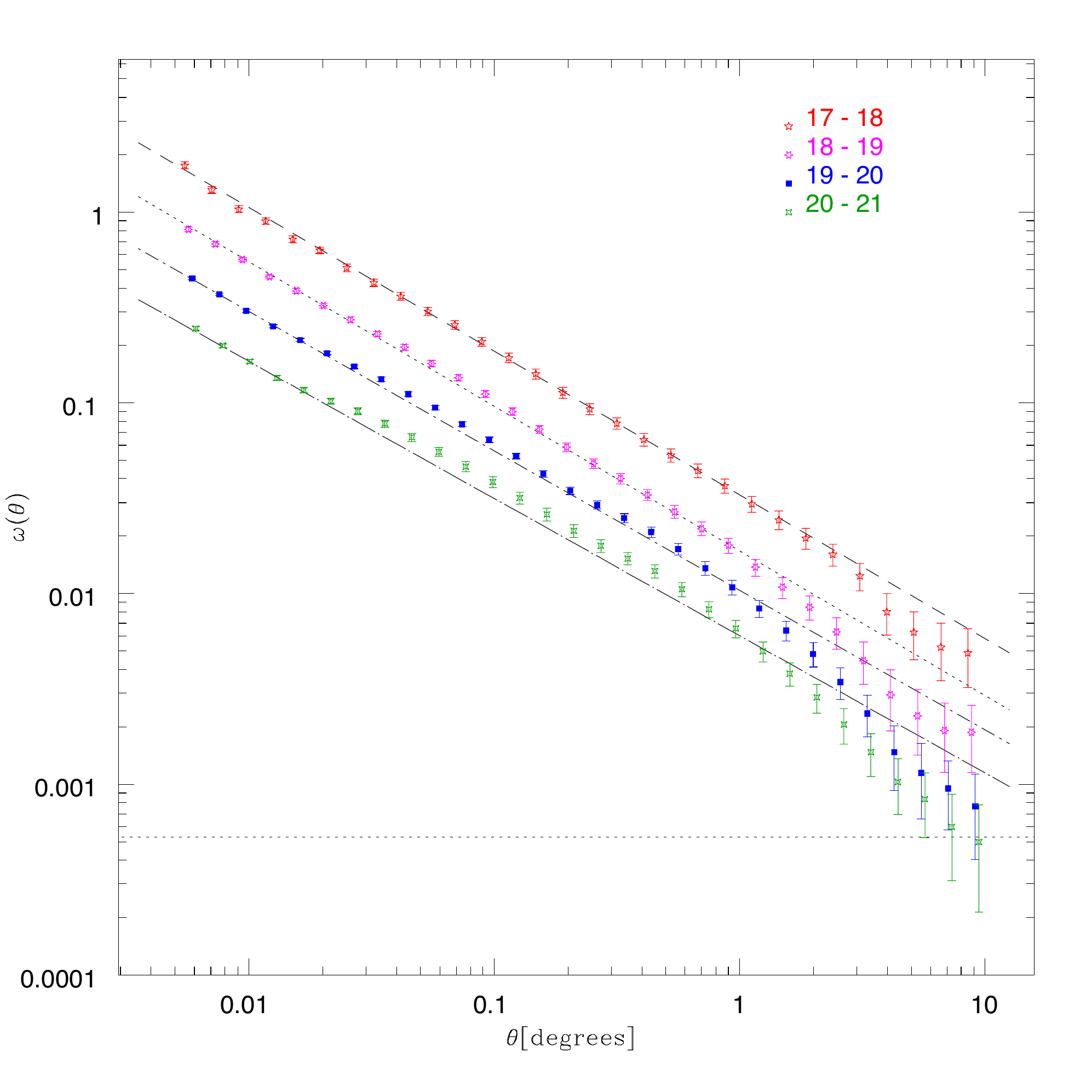}}}
\end{center}
\caption{\B{Left: The two-point  angular correlation function for galaxies in the magnitude range $17 < r \le 21$. Right: The two-point galaxy angular correlation function split by magnitude: $17 < r \le18$ (red), $18 < r \le 19$ (magenta), $19 < r \le 20$ (blue), and $20 < r \le 21$ (green). Overplotted for all four correlation function measurements are the best fit power laws: $\omega(\theta) = A_\omega \theta^{(1-\gamma)}$, the individual fit values: $A_{\omega}$ and$\gamma$ are given in Table~\ref{17to21magTypes}. In both plots we draw a line at $\omega(\theta) = 0.00053$, which is the typical scale of the maximum systematic contamination (\ie Galactic extinction ) at large angles (see, \eg Figures~\ref{cross17to21} and~\ref{galcross}).}}
\label{17to21all}
\end{figure*}

Fundamentally, the calculation of the 2PCF involves determining how many pairs of data points lie within particular distance bins as compared to a Poisson distribution. For a data set with $N$ points, the naive approach calls for the calculation of the distance from each point to all other $N - 1$ points. Clearly, this approach leads to a computational load that scales as $O(N^2)$, which proves impractical for large $N$. One can improve the calculation by organizing the data into a two-dimensional quad-tree structure that groups nearby data points~\citep{Moore01}. This technique gives significant savings in practice, as instead of calculating the distance to every point, one can often account for entire groups of points by looking only at the bounding boxes of the different groups.

We first perform preliminary calculations for all galaxies and random points that convert and organize the data into an optimized format for subsequent calculations. The pre-computing codes are all serial, as this step in the 2PCF calculation is relatively inexpensive. Next, we construct the quad-tree for both the galaxies and random points by using a modified $k$d-tree~\citep{Bent}. This produces a balanced tree with minimal depth that both minimizes the memory required to store the tree and leads to an efficient tree traversal in later computations. 

The algorithm proceeds by computing the minimum size bounding box that contains all the data (\ie root node), which is subsequently subdivided in a recursive manner into new subsamples (\ie child nodes). We quantify the minimum size bounding box for each subsample, and this process continues until a child node contains fewer data points than a preset limit. We also consider the jackknife resampling when building the trees since we must be sure to select data and random points that occupy the same volume for a given sample for each jackknife.

Given the tree structure above, we must be able to quickly determine the minimum and maximum angular separations between two nodes or a point and a node. Since we have stored the $cosine$ and $sine$ of the angular size of each node as well as their centers, the requisite information can be computed without ever using a trigonometric function evaluation.

To parallelize this algorithm, we note that the comparison of two data sets represented by their trees can be broken into subproblems by comparing all nodes at a given level $L$ in one tree with the root node of the other. Since we use binary trees, this yields $2^{L}$ subproblems that can be distributed to multiple processors. We employ a master-slave arrangement where one processor is responsible for coordinating the parallel calculation and the remaining processors make requests for work as needed. When not handling work requests the master process performs smaller amounts of work by descending deeper into the tree, which assures that it frequently checks for work requests but is not idle when no requests have been posted. In the current implementation, all processors have direct access to all the data which limits communication to single integer tags identifying particular subproblems.

\subsubsection{\em The Performance}

To demonstrate the performance and scaling of this implementation, we ran each correlation function ten times and compute the mean time and the standard deviation of the ten separate calculations. Figure~\ref{timescale} shows how the code scales with an increasing number of galaxies by using only one processor (left panel) and with an increasing number of processors by using two million galaxies (right panel). For each galaxy sample, we use random data with ten times more points in the same sky region and we compute the angular correlations to $10\degr$. The left plot in Figure~\ref{timescale} shows that the runtime scales with $N^a$, where $a\simeq1.35$. The right plot shows how the running time scales with number of processors, with 2 million points in the galaxy sample. These two plots illustrate that the parallel algorithm we present above computes the 2PCF efficiently over a wide range of angles for large data sets.

\subsection{\em The Angular Correlation Function of DR7 Galaxies}

\begin{table*}
\begin{center}
\caption{The two-point galaxy angular correlation function measurements for the full galaxy sample and our four magnitude limited sub-samples: $17 < r \leq 18$, $18 < r \leq 19$, $19 < r \leq 20$, and $20 < r \leq 21$)}
{\begin{tabular}{c c c c c c}\\ \hline \hline
Angle(deg) & $17 < r \leq 21$ & $17 < r \leq 18$ & $18 < r \leq 19$ & $19 < r \leq 20$ & $20 < r \leq 21$ \\ \hline
0.006 & 0.3173 $\pm$ 0.0045 & 1.7583 $\pm$ 0.0797 & 0.8129 $\pm$ 0.0220 & 0.4498 $\pm$ 0.0074 & 0.2456 $\pm$ 0.0049\\
0.007 & 0.2548 $\pm$ 0.0040 & 1.3099 $\pm$ 0.0559 & 0.6805 $\pm$ 0.0182 & 0.3705 $\pm$ 0.0059 & 0.1995 $\pm$ 0.0047\\
0.009 & 0.2093 $\pm$ 0.0034 & 1.0344 $\pm$ 0.0448 & 0.5634 $\pm$ 0.0149 & 0.3041 $\pm$ 0.0055 & 0.1646 $\pm$ 0.0043\\
0.012 & 0.1718 $\pm$ 0.0031 & 0.8975 $\pm$ 0.0352 & 0.4582 $\pm$ 0.0112 & 0.2520 $\pm$ 0.0037 & 0.1349 $\pm$ 0.0040\\
0.016 & 0.1458 $\pm$ 0.0029 & 0.7198 $\pm$ 0.0325 & 0.3871 $\pm$ 0.0113 & 0.2134 $\pm$ 0.0037 & 0.1166 $\pm$ 0.0038\\
0.020 & 0.1269 $\pm$ 0.0027 & 0.6310 $\pm$ 0.0260 & 0.3241 $\pm$ 0.0087 & 0.1824 $\pm$ 0.0038 & 0.1022 $\pm$ 0.0039\\
0.026 & 0.1106 $\pm$ 0.0025 & 0.5108 $\pm$ 0.0245 & 0.2733 $\pm$ 0.0093 & 0.1549 $\pm$ 0.0034 & 0.0904 $\pm$ 0.0036\\
0.033 & 0.0943 $\pm$ 0.0023 & 0.4274 $\pm$ 0.0193 & 0.2296 $\pm$ 0.0067 & 0.1328 $\pm$ 0.0032 & 0.0779 $\pm$ 0.0034\\
0.043 & 0.0794 $\pm$ 0.0021 & 0.3619 $\pm$ 0.0183 & 0.1957 $\pm$ 0.0068 & 0.1112 $\pm$ 0.0028 & 0.0660 $\pm$ 0.0032\\
0.056 & 0.0664 $\pm$ 0.0020 & 0.3019 $\pm$ 0.0146 & 0.1611 $\pm$ 0.0061 & 0.0942 $\pm$ 0.0026 & 0.0554 $\pm$ 0.0030\\
0.072 & 0.0552 $\pm$ 0.0018 & 0.2564 $\pm$ 0.0139 & 0.1385 $\pm$ 0.0052 & 0.0773 $\pm$ 0.0023 & 0.0463 $\pm$ 0.0028\\
0.092 & 0.0455 $\pm$ 0.0017 & 0.2091 $\pm$ 0.0114 & 0.1114 $\pm$ 0.0046 & 0.0642 $\pm$ 0.0021 & 0.0384 $\pm$ 0.0025\\
0.119 & 0.0375 $\pm$ 0.0015 & 0.1723 $\pm$ 0.0102 & 0.0899 $\pm$ 0.0041 & 0.0525 $\pm$ 0.0018 & 0.0318 $\pm$ 0.0023\\
0.153 & 0.0306 $\pm$ 0.0013 & 0.1419 $\pm$ 0.0087 & 0.0726 $\pm$ 0.0035 & 0.0425 $\pm$ 0.0016 & 0.0260 $\pm$ 0.0020\\
0.197 & 0.0250 $\pm$ 0.0012 & 0.1133 $\pm$ 0.0074 & 0.0585 $\pm$ 0.0030 & 0.0346 $\pm$ 0.0015 & 0.0213 $\pm$ 0.0017\\
0.254 & 0.0209 $\pm$ 0.0011 & 0.0929 $\pm$ 0.0061 & 0.0480 $\pm$ 0.0027 & 0.0291 $\pm$ 0.0014 & 0.0178 $\pm$ 0.0014\\
0.327 & 0.0178 $\pm$ 0.0011 & 0.0781 $\pm$ 0.0054 & 0.0400 $\pm$ 0.0025 & 0.0249 $\pm$ 0.0014 & 0.0153 $\pm$ 0.0011\\
0.421 & 0.0152 $\pm$ 0.0011 & 0.0642 $\pm$ 0.0049 & 0.0330 $\pm$ 0.0022 & 0.0210 $\pm$ 0.0013 & 0.0131 $\pm$ 0.0010\\
0.543 & 0.0123 $\pm$ 0.0010 & 0.0531 $\pm$ 0.0042 & 0.0268 $\pm$ 0.0021 & 0.0171 $\pm$ 0.0012 & 0.0105 $\pm$ 0.0009\\
0.699 & 0.0097 $\pm$ 0.0009 & 0.0440 $\pm$ 0.0036 & 0.0219 $\pm$ 0.0018 & 0.0136 $\pm$ 0.0011 & 0.0083 $\pm$ 0.0008\\
0.901 & 0.0078 $\pm$ 0.0008 & 0.0367 $\pm$ 0.0032 & 0.0179 $\pm$ 0.0016 & 0.0108 $\pm$ 0.0010 & 0.0066 $\pm$ 0.0007\\
1.161 & 0.0060 $\pm$ 0.0007 & 0.0294 $\pm$ 0.0029 & 0.0137 $\pm$ 0.0014 & 0.0084 $\pm$ 0.0008 & 0.0050 $\pm$ 0.0006\\
1.495 & 0.0046 $\pm$ 0.0006 & 0.0243 $\pm$ 0.0027 & 0.0108 $\pm$ 0.0014 & 0.0064 $\pm$ 0.0007 & 0.0038 $\pm$ 0.0005\\
1.927 & 0.0035 $\pm$ 0.0006 & 0.0195 $\pm$ 0.0024 & 0.0085 $\pm$ 0.0012 & 0.0048 $\pm$ 0.0007 & 0.0029 $\pm$ 0.0005\\
2.482 & 0.0026 $\pm$ 0.0005 & 0.0160 $\pm$ 0.0021 & 0.0063 $\pm$ 0.0012 & 0.0034 $\pm$ 0.0006 & 0.0021 $\pm$ 0.0004\\
3.198 & 0.0018 $\pm$ 0.0005 & 0.0124 $\pm$ 0.0020 & 0.0045 $\pm$ 0.0011 & 0.0023 $\pm$ 0.0006 & 0.0015 $\pm$ 0.0004\\
4.120 & 0.0012 $\pm$ 0.0004 & 0.0080 $\pm$ 0.0020 & 0.0029 $\pm$ 0.0010 & 0.0015 $\pm$ 0.0006 & 0.0010 $\pm$ 0.0003\\
5.308 & 0.0010 $\pm$ 0.0004 & 0.0063 $\pm$ 0.0018 & 0.0023 $\pm$ 0.0009 & 0.0011 $\pm$ 0.0005 & 0.0008 $\pm$ 0.0003\\
6.838 & 0.0007 $\pm$ 0.0003 & 0.0052 $\pm$ 0.0017 & 0.0019 $\pm$ 0.0008 & 0.0010 $\pm$ 0.0004 & 0.0006 $\pm$ 0.0003\\
8.810 & 0.0006 $\pm$ 0.0003 & 0.0049 $\pm$ 0.0017 & 0.0019 $\pm$ 0.0007 & 0.0008 $\pm$ 0.0004 & 0.0005 $\pm$ 0.0003\\ \hline
\label{17to21magCorr}
\end{tabular}}
\end{center}
\end{table*}

\begin{table}
\begin{center}
\caption{Parameter values for the power-law model fits, for both the full galaxy sample and magnitude limited subsamples.)}
{\begin{tabular}{c c c c c }\\ \hline \hline
Magnitude & $log_{10}A_{\omega}$ & $1-\gamma$ & $\chi^2$/dof \\ \hline \hline
$17 < r \leq 21$ & & & \\
$ (full\ sample) $ & -2.120 $\pm$ 0.019 & -0.720 $\pm$ 0.010 & 5.30\\ \hline
$17 < r \leq 18$ & -1.483 $\pm$ 0.009 & -0.754 $\pm$ 0.006 & 0.76\\
$18 < r \leq 19$ & -1.776 $\pm$ 0.014 & -0.759 $\pm$ 0.008 & 2.36\\
$19 < r \leq 20$ & -1.983 $\pm$ 0.018 & -0.731 $\pm$ 0.010 & 5.46\\
$20 < r \leq 21$ & -2.222 $\pm$ 0.023 & -0.719 $\pm$ 0.012 & 4.26\\ \hline \hline
\label{17to21magTypes}
\end{tabular}}
\end{center}
\end{table}

\B{By applying the correlation function estimator in Equation~\ref{landEq} to the full galaxy sample, as defined by the restrictions outlined in Section~\ref{SysResults}, we measure the two-point galaxy angular correlation function for the SDSS DR7. The resulting correlation function is shown in the left-hand panel of Figure~\ref{17to21all}. By calculating the full sample correlation matrix estimator shown in Equation~\ref{covmat}, we  obtain a model power law fit using Equation~\ref{curfit}, finding $A_{\omega} = -2.12$ with $\gamma \simeq 1.72$ for the full galaxy sample, which is over plotted with the data shown in Figure~\ref{17to21all}. We present the full sample correlation matrix, which is computed from the covariance matrix~\citep[see, \eg][]{Scr}:}
\begin{equation}
r(\theta_{i}, \theta_{j}) = \frac{C(\theta_{i}, \theta_{j})}{\sqrt{C(\theta_{i}, \theta_{i}) C(\theta_{j}, \theta_{j})}}
\label{cormat}
\end{equation}
The correlation matrix for the full galaxy sample is presented in Figure~\ref{cov}, and is seen to be highly diagonal. We tabulate all covariance matrices in Appendix~\ref{covtab}. Overall, the amplitude of the correlation function is consistent with previous results from surveys such as the APM~\citep{Mad} and the SDSS EDR~\citep{Con}; a more detailed comparison is presented in Section~\ref{Conclusion}.

\begin{figure}
\begin{center}
{\resizebox{8 cm}{!}{\includegraphics{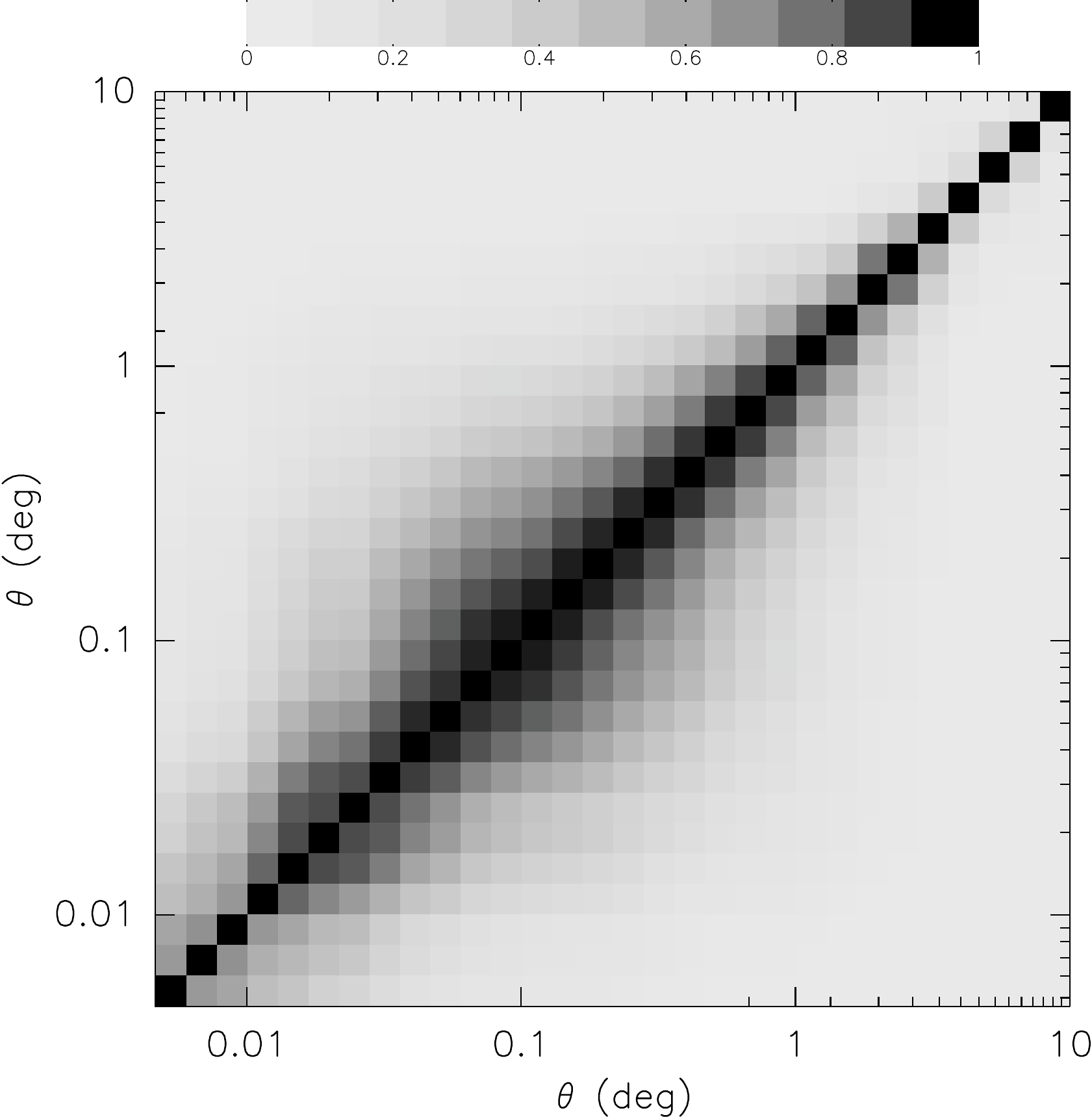}}}
\end{center}
\caption{{The correlation matrix for the full galaxy sample, pixel values range from $0$ (uncorrelated; white) to $1$ (fully correlated; black).}}
\label{cov}
\end{figure}

By following this same procedure, we measure the two-point galaxy angular correlation function for four magnitude limited samples: $17 < r \leq 18$, $18 < r \leq 19$, $19 < r \leq 20$, and $20 < r \leq 21$, which are shown in the right-hand panel of Figure~\ref{17to21all}. The actual two-point galaxy angular correlation measurements are also presented in Table~\ref{17to21magCorr} for each angular bin.

We fit these individual angular correlation functions, following the same technique as described for the full galaxy sample, but by using the appropriate magnitude range jackknife covariance matrix. The best fit power-law models are overplotted on the relevant data in Figure~\ref{17to21all}, and the power-law fit parameters are tabulated for the full galaxy sample and each of the four magnitude limited samples in Table~\ref{17to21magCorr}. We find that the amplitudes of these four correlation functions are found to decrease with increasing magnitude, as expected since we are sampling intrinsically fainter galaxies that are known to be clustered less strongly~\citep{Scr,Con}.

\section{Discussions and Conclusions}\label{Conclusion}

In this paper, we present the first, complete measurement of the two-point galaxy angular correlation function for the SDSS DR7. 
To make the most precise measurement possible, we first perform a thorough reanalysis of possible external and internal sources of error. First, we found that the SDSS DR7 data have a detection completeness of approximately 90\% to a dereddended $r$-band model magnitude of $21$. Second, we demonstrated that the source classification is 95\% complete to this same magnitude limit. Thus we restrict dour galaxy sample to have $r$-band magnitudes in the range 17 $< r \leq$ 21. Next, we confirmed the overall quality of the SDSS photometric data, and find that our signal is maximized by restricting the SDSS DR7 data to those regions of the survey that have seeing $< 1\farcs5$ and reddening $< 0.13$ mag. With these sample restrictions, the majority of the data from stripe 42, 43 and 44 are removed; therefore, for simplicity we simply exclude these three stripes from final galaxy sample. 

\B{We also explored the effect of the SDSS survey strategy on our measurement, finding that the variations in galaxy densities and $\omega(\theta)$ across stripes are small,. Therefore, we see no reason to {\it a priori} exclude any of the remaining thirty-four stripes that constitute our final sample. Finally, we compared the two-point galaxy auto-correlation function with the two-point cross-correlation function between galaxies and seeing, and between galaxies and reddening, finding that the amplitudes of these systematic errors are well below the measurement of $\omega(\theta)$ on angular scales from $0\fdg05$ to $5\degr$. From these measured systematics, we can suggest that, unless these systematic effects can be mitigated more effectively,  the measurement of galaxy angular correlation functions  from forthcoming large surveys, such as DES and LSST, will be limited to smaller angular scales as they will probe intrinsically fainter magnitudes. One method to mitigate this effect, however, will be to use photometric redshifts to divide the angular signal into smaller redshift shells to minimize the projection effects in measuring $\omega(\theta)$ and thereby increase the amplitude of the overall signal.}

One result of our analysis of different systematics was that stars do not play a major effect in the SDSS DR7. This is in direct conflict with the results of~\citet{R11}, who demonstrated that stars are one of the dominant contaminants in clustering measurements of luminous red galaxies in the SDSS DR8. These differences can be explained by several facts. First, we explore the effects of the different systematics on all galaxies, not just luminous red galaxies, which are generally quite faint in the SDSS imaging data. Second, SDSS DR7 does not extend to the same low Galactic latitudes as SDSS DR8, which means DR8 will include regions of much higher stellar density. Third, we use more stringent swing and reddening cuts, which will reduce the overall sky coverage, preferentially to higher Galactic latitudes. Fourth, we use the official SDSS star/galaxy classification method, as opposed to the a separate neural-network classification. Finally, the SDSS DR8 has a known photometric issue that affects the measured color offsets as copared to the SDSS DR7 photometric pipeline~\citep{R11}.

\begin{figure}
\begin{center}
{\resizebox{8 cm}{!}{\includegraphics{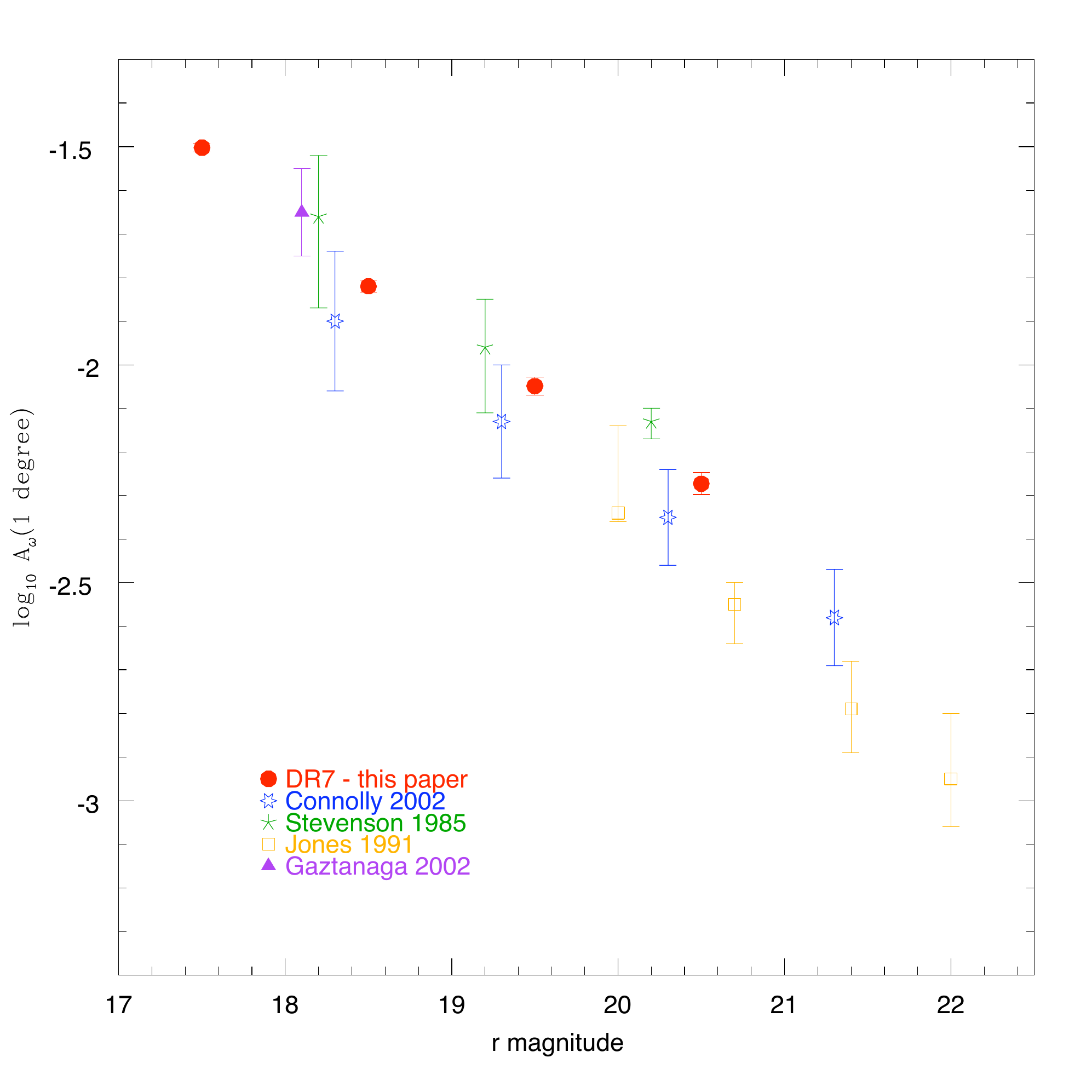}}}
{\resizebox{8 cm}{!}{\includegraphics{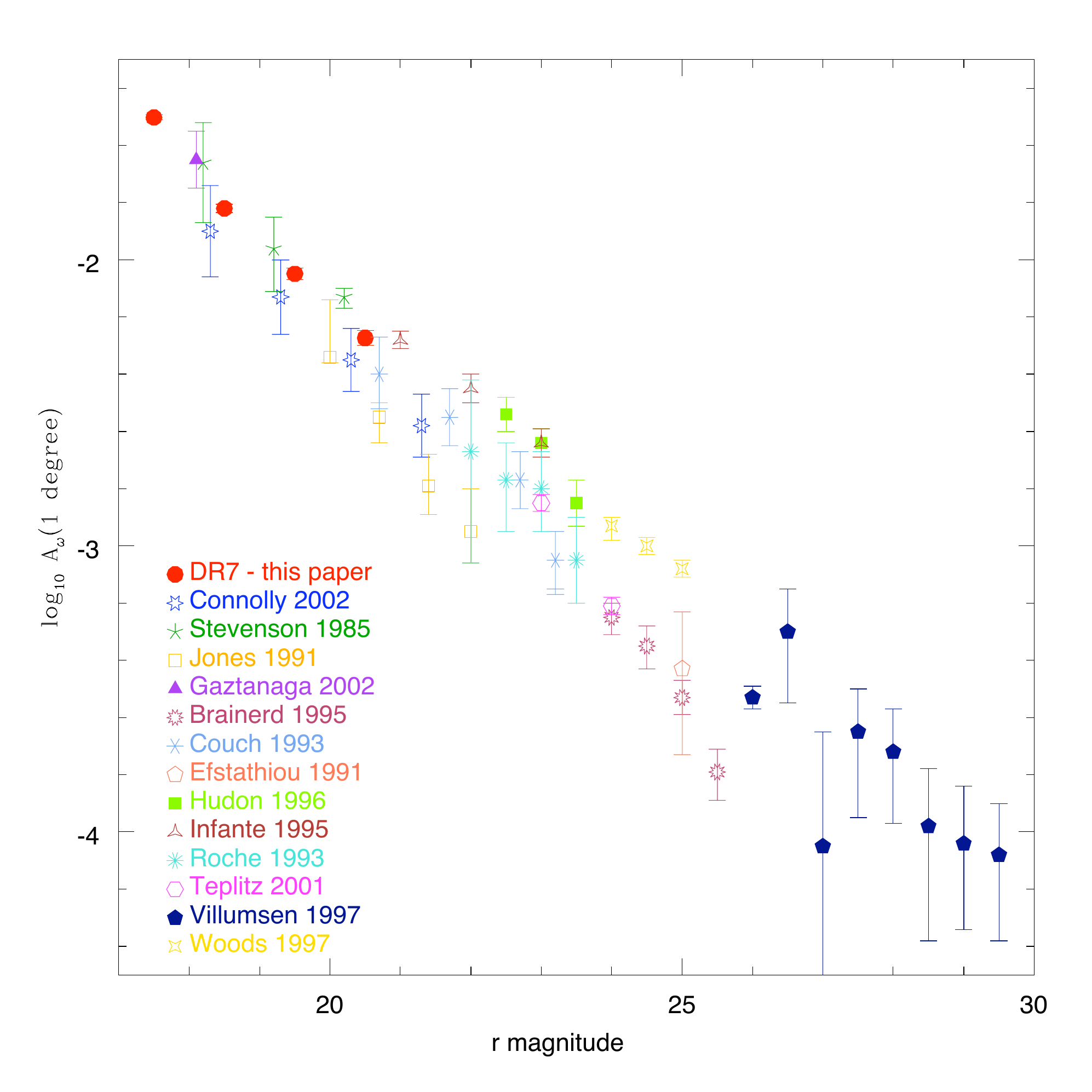}}}
\end{center}
\caption{{The correlation function amplitude versus $r$-band magnitude at $\theta=1\degr$. Top: The comparison of our result with the results from~\citet{Con} and~\citet{Gazt}, we shift these two results by $0.2$ magnitudes to account for the known SDSS EDR photometry problem, and~\citet{Steve}. Bottom: The comparison of these correlation amplitudes to correlation amplitudes measured from fainter galaxy catalogs (see text for details). Note that we have ignored the small differences between the various r-band filters used by these different authors.}}
\label{comfig}
\end{figure}

\B{Our final measurement of the two-point galaxy angular correlation function includes data taken through August 2008, and demonstrates that both the shape and amplitude of the two-point galaxy angular correlation function are similar to (albeit more precise than) previous published results. We find that the correlation function can generally be described by a power law $\omega(\theta) = A_\omega \theta^{(1-\gamma)}$, with $\gamma \simeq 1.72$ on both small and large scales. The amplitude of the correlation function decreases as a function of magnitude, which is also in good agreement with previous results, with $\gamma \simeq$ $1.75$, $1.76$, $1.73$, and $1.72$ for magnitude bins $17 < r \leq 18$, $18 < r \leq 19$, $19 < r \leq 20$, and $20 < r \leq 21$.}

In Figure~\ref{comfig}, we compare our galaxy angular correlation function amplitudes at $\theta=1\degr$ for magnitude bins $17 < r \leq 18$, $18 < r \leq 19$, $19 < r \leq 20$, and $20 < r \leq 21$ with previous, published results made from other galaxy catalogs (note that we have made no effort to correct for the likely small, and unknown differences in the various r-band filters used by the different authors). At brighter magnitudes these catalogs include the SDSS EDR~\citep[Sloan Digital Sky Survey Early Data Release:][]{Con,Gazt} and UKST~\citep[UK Schmidt Telescope:][]{Steve}. While at fainter magnitudes, we compare  to  galaxy catalogs from the AAT~\citep[Anglo-Australian Telescope:][]{Jones,Couch}, Hale Telescope~\citep{Brain}, TS12 (\citet{Efst}), CFHT~\citep[Canada-France-Hawaii Telescope:][]{Hudon,Woods,Inf95}, INT~\citep[Isaac Newton Telescope:][]{Roche}, the HDF~\citep[Hubble Deep Field:][]{Vill}; and the HDF-South~\citep{Tep}. 

Overall, our measurement is quite precise, which is expected given the uniformity of our data and the extremely large size of our galaxy sample. The top panel in Figure~\ref{comfig} shows the comparison with previous efforts that cover the same magnitude range as our data. For this figure, we have shifted the EDR measurements (\citet{Con} and \citet{Gazt}) by 0.2 magnitudes to account for the known SDSS EDR photometry error~\citep{Aba04}. In general, our amplitudes agree well with \citet{Steve}, \citet{Gazt} and \citet{Jones}. While our measured clustering strength is within one-sigma of the SDSS EDR results of~\citet{Con}, they are not as close as one might think. However, we have shown that both the galaxy density and galaxy clustering strength are mildly stripe dependent (see, \eg Figures~\ref{dataDen} and~\ref{stripesfig}), thus it is not surprising that there would be differences between the single EDR stripe and our full, thirty-four stripe DR7 sample. Finally, we note that our results agree with the general trend shown by previous results at fainter magnitudes.

While we have presented the first, complete SDSS galaxy angular correlation measurement in this paper, there remains considerable work to do in this area. First, our analysis of the resulting correlation functions in this paper has followed the standard power-law clustering model~\citep[\eg][]{Brunner00, Tep}. These models are no longer as popular, in part since they do not fully capture the full nuances of the galaxy angular correlation function as measured from large, uniform data sets. We see this in the reduced $\chi^2$ values for our fits, which are $4.26$ for the faintest magnitude function and $5.30$ for the full sample. Newer approaches have been developed~\citep[\eg][]{Brown08, Coupon12} that allow stronger constraints to be placed on structure formation models. In addition, halo models~\cite[\eg][]{Zheng05} have been developed for the interpretation of galaxy spatial clustering measurements, and these models have been extended to the analysis of angular clustering measurements when augmented by photometric redshifts~\citep{R06, Ross07}. Thus, an interesting next step will be to extend our measurements presented in this paper to use the SDSS photometric redshift estimates~\citep{Csabai07} within these more advanced techniques~\citep{Coupon12}.

In addition, the SDSS photometric redshift estimation process also provides a spectral type classification that can be used to divide the SDSS galaxy sample into early- and late-type galaxy samples~\citep[see, \eg][]{Budavari03}. Previous efforts have used these type classifications along with the photometric redshift estimates to construct volume-limited galaxy samples (that can also be further subdivided by galaxy type) to measure the evolution of the angular clustering of a volume-limited sample of galaxies via correlation functions~\citep{Ross09, Ross10} and via the angular power spectrum~\citep{Hayes12}. While this can easily be done with our current sample as a consistency check, a more interesting analysis would be to find photometrically classified galaxy pairs, triplets, and quads to explore their clustering behavior both in general and as a function of galaxy type and redshift. This would produce a new approach to the study of central and satellite galaxy distributions within halo occupation distribution models.

\section*{Acknowledgements}

The authors would like to thank Brett Hayes, Adam Myers,  Ashley Ross, and the anonymous referee for valuable discussion and advice that have greatly improved this paper.

Funding for the SDSS and SDSS-II has been provided by the Alfred P. Sloan Foundation, the Participating Institutions, the National Science Foundation, the U.S. Department of Energy, the National Aeronautics and Space Administration, the Japanese Monbukagakusho, the Max Planck Society, and the Higher Education Funding Council for England. The SDSS Web Site is http://www.sdss.org/.

The Sloan Digital Sky Survey (SDSS) is a joint project of the american Museum of Natural History, Astrophysical Institute Potsdam, University of Basel, University of Cambridge, Case Western Reserve University, University of Chicago, Drexel University, Fermilab, the Institute for Advanced Study, the Japan Participation Group, Johns Hopkins University, the Joint Institute for Nuclear Astrophysics, the Kavli Institute for Particle Astrophysics and Cosmology, the Korean Scientist Group, the Chinese Academy of Sciences (LAMOST), Los Alamos National Laboratory, the Max-Planck-Institute for Astronomy (MPIA), the Max-Planck-Institute for Astrophysics (MPA), New Mexico State University, Ohio State University, University of Pittsburgh, University of Portsmouth, Princeton University, the United States Naval Observatory, the University of Washington.

\bibliographystyle{mn2e}
\bibliography{2012Mar}

\appendix

\section{Creating the galaxy sample}\label{AppCuts}
In this Appendix, we detail the steps we take to obtain our final galaxy catalog used in the analysis presented herein, starting with our original SQL query to the SDSS catalog archive server.

\subsection{The SDSS CAS SQL Query}\label{query}
Our first step is to extract all relevant data from the SDSS catalog archive server. We do this by issuing the following SQL query:

\begin{alltt}
SELECT

    p.objID, p.ra, p.dec, 
    p.type, p.flags, p.insideMask,

-- -- First get PSF Mags

    p.psfMag\_u, p.psfMagErr\_u, 
    p.psfMag\_g, p.psfMagErr\_g, 
    p.psfMag\_r, p.psfMagErr\_r, 
    p.psfMag\_i, p.psfMagErr\_i, 
    p.psfMag\_z, p.psfMagErr\_z, 

-- -- Now get Model Mags

    p.modelMag\_u, p.modelMagErr\_u, 
    p.modelMag\_g, p.modelMagErr\_g, 
    p.modelMag\_r, p.modelMagErr\_r, 
    p.modelMag\_i, p.modelMagErr\_i, 
    p.modelMag\_z, p.modelMagErr\_z, 

-- -- Now get Petro Mags

    p.petroMag\_u, p.petroMagErr\_u, 
    p.petroMag\_g, p.petroMagErr\_g, 
    p.petroMag\_r, p.petroMagErr\_r, 
    p.petroMag\_i, p.petroMagErr\_i, 
    p.petroMag\_z, p.petroMagErr\_z, 

-- -- Now get Fiber Mags

    p.fiberMag\_u, p.fiberMagErr\_u, 
    p.fiberMag\_g, p.fiberMagErr\_g, 
    p.fiberMag\_r, p.fiberMagErr\_r, 
    p.fiberMag\_i, p.fiberMagErr\_i, 
    p.fiberMag\_z, p.fiberMagErr\_z, 

-- -- Get concentration parameters

    p.petroR50\_r, p.petroR90\_r, 

-- -- Get all extinction values

    p.extinction\_u, p.extinction\_g, 
    p.extinction\_r, p.extinction\_i, 
    p.extinction\_z, 

-- -- Get all type and flag information

    p.type\_u, p.type\_g, p.type\_r, 
    p.type\_i, p.type\_z, 
    
    p.flags\_u, p.flags\_g, p.flags\_r, 
    p.flags\_i, p.flags\_z, 

-- -- Get Michigan Moments for seeing

    p.mRrCc\_u, p.mRrCcErr\_u, p.mRrCcPSF\_u, 
    p.mRrCc\_g, p.mRrCcErr\_g, p.mRrCcPSF\_g, 
    p.mRrCc\_r, p.mRrCcErr\_r, p.mRrCcPSF\_r, 
    p.mRrCc\_i, p.mRrCcErr\_i, p.mRrCcPSF\_i, 
    p.mRrCc\_z, p.mRrCcErr\_z, p.mRrCcPSF\_z, 

-- -- Now get all photoz values

    z.z, z.zErr, z.chiSq, z.nnIsInside, z.pztype, 
    z.dmod, z.kcorr\_u, z.kcorr\_g, z.kcorr\_r, 
    z.kcorr\_i, z.kcorr\_z, z.absMag\_u, z.absMag\_g, 
    z.absMag\_r, z.absMag\_i, z.absMag\_z, 

-- -- Now the Table join

FROM  PhotoPrimary  AS  p 

LEFT OUTER JOIN
    photoz AS z ON p.objID = z.objID 

-- -- Limit output to reasonable detections

WHERE
    ((p.dered\_g < 23.0) OR 
     (p.dered\_r < 23.0) OR 
     (p.dered\_i < 23.0))
\end{alltt}
This request generates a catalog with more than $341$ million point sources, including stars and galaxies.

\subsection{Cutting to the SDSS Theoretical Footprint}\label{footprint}

The SDSS DR7 footprint is defined by all non-repeating, survey-quality imaging runs observed prior to July 2008, including the elliptical survey area in the Northern Hemisphere and the three stripes in the Southern Hemisphere. Starting with the catalog returned by our SQL query, we first apply the SDSS theoretical footprint\footnote{\url{http://www.sdss.org/dr7/coverage/index.html}}. From this cut, we retain thirty-one stripes in the Northern Hemisphere: $9$--$39$, and three stripes from the Southern Hemisphere: $76$, $82$, $86$. After these cuts, our catalog has $\sim214$ million sources (\ie we keep 62.8\% of the data from the original catalog), covering $\sim8200$ square degrees of the sky, with $\sim7650$ square degrees of this in the Northern Galactic Cap high-latitude region and $\sim750$ square degrees of the total from the three stripes in the Southern Galactic Cap.

\AC{Furthermore, we mask objects that are in any of the five image masks~\footnote{\url{http://www.sdss.org/dr7/algorithms/masks.html}}, and this keeps 94.6\% of the above data, which results in 203 million objects.}

\subsection{Applying object detection and measurement flags}\label{flagCut}
For completeness, we detail the relevant SDSS photometric flags in Table~\ref{flags}. In this section, we outline the method by which we follow the SDSS project recommendations\footnote{\url{http://www.sdss.org/dr7/products/catalogs/flags.html}} to restrict our sample to clean, photometric detections by using the object flags assigned by the SDSS photometric pipeline. First we compute the following two meta-flags:

\begin{alltt}
DEBLEND\_PROBLEMS =
    PEAKCENTER || 
    NOTCHECKED ||
    (DEBLEND\_NOPEAK \&\& psfErr\_r > 0.2)

INTERP\_PROBLEMS = 
    PSF\_FLUX\_INTERP || 
    BAD\_COUNTS\_ERR || 
    (INTERP\_CENTER \&\& CR)
\end{alltt}

which simplifies subsequent flag tests. For \texttt{DEBLEND\_PROBLEMS}, if either \texttt{PEAKCENTER} or \texttt{NOTCHECKED} is set, or if \texttt{psfErr\_r} is greater than $0.2$ magnitudes and \texttt{DEBLEND\_NOPEAK} is set, we set the \texttt{DEBLEND\_PROBLEMS} meta-flag. Likewise for  \texttt{INTERP\_PROBLEMS}, if the\texttt{PSF\_FLUX\_INTERP} or \texttt{BAD\_COUNTS\_ERROR} flags are set, or if \texttt{INTERP\_CENTER} and \texttt{CR} are both set, we set the \texttt{INTERP\_PROBLEMS} meta-flag. In the end, we only accept objects that pass the following, r-band flag test:
\begin{alltt}
\AC{BINNED1 \&\& !SATURATED \&\& !NOPROFILE \&\& }
\AC{!INTERP\_PROBLEMS \&\& !DEBLEND\_PROBLEMS \&\&}
\AC{(!EDGE || }
\AC{    (EDGE \&\& (NODEBLEND || DEBLENDED\_AT\_EDGE))) \&\& }
\AC{!(DEBLENDED_AS_PSF \&\& }
\AC{    (TOO_FEW_GOOD_DETECTION || NOPETRO))}
\end{alltt}

\begin{table*}
\caption{This table briefly describes all flags that may affect the quality of the SDSSS imaging data. The percentage of data with each flag is based on the SQL query in~\ref{query}. The flags that have and asterisk (*) appended can be set in single band.}
{\begin{tabular}{r l c}\\ \hline \hline
Name & Description & Data with \\ 
& & this\ flag\ (\%) \\ \hline \hline
\multicolumn{2}{| l }{ \texttt{INTERP\_PROBLEMS}: } \\ \cline{1-1}
\texttt{PSF\_FLUX\_INTERP}* & More than 20\% of the PSF flux is from interpolated pixels. & 14.7 \\ 
 \texttt{BAD\_COUNTS\_ERR}* & The object contains many interpolation affected pixels, thus there are too few  & 0\\
& good pixels to estimate a PSF error. \\ 
 \texttt{INTERP\_CENTER} & The interpolated pixel is within 3 pixels of the object center. & 9.28\\
 \texttt{COSMIC\_RAY (CR)} & The object contains cosmic rays. & 12.3 \\ \hline \hline

\multicolumn{2}{| l }{\texttt{DEBLEND\_PROBLEMS}: } \\ \cline{1-1}
\texttt{PEAKCENTER}* & The object uses the position of the peak pixel as its center. & 0.549\\
\texttt{NOTCHECKED} & The object contains pixels that were not checked to see if they include local peaks. & 1.20\\
\texttt{DEBLEND\_NOPEAK} & The object is a \texttt{CHILD} but has no peak in at least one band. & 11.9\\
\texttt{psfErr\_r} & PSF flux error in r-band. & 28.2 \\ \hline \hline 

\multicolumn{2}{| l }{\emph{Object status flags}: } \\ \cline{1-1}
\texttt{BINNED1}* & The object was detected at $\ge 5 \sigma$ in a $1\times1$ binned image. & 97.2\\
\texttt{BINNED2}* & The object was detected in a $2\times2$ binned image. & 2.97\\
\texttt{BINNED4}* & The object was detected in a $4\times4$ binned image. & 0.105\\
\texttt{DETECTED} & The object was either detected in \texttt{BINNED1}, \texttt{BINNED2}, or \texttt{BINNED4}. & 99.8\\
\texttt{BRIGHT} & The object was duplicate-detected at $> 200 \sigma$, which usually means $r < 17.5$. & 0 \\
\texttt{BLENDED}* & The object was detected with multiple peaks, and thus there was an attempt to deblend & 9.15 \\
&  it as a parent object. & \\
\texttt{NODEBLEND} & The object was \texttt{BLENDED}, but there was no attempt to deblend it. & \\
\texttt{CHILD} & The object was the result of deblending a \texttt{BLENDED} object. It may still be \texttt{BLENDED}. & 26.1\\ \hline \hline

\multicolumn{2}{| l }{\emph{Raw data problem flags}: } \\ \cline{1-1}
\texttt{SATURATED}* & The object contains one or more saturated pixels. & 4.42\\
\texttt{EDGE} & The object is too close to the edge of a field frame. & 0.432\\
\texttt{LOCAL\_EDGE} & Similar to \texttt{EDGE}, but one half of the CCD failed. & 0\\
\texttt{DEBLENDED\_AT\_EDGE} & The object is so large that it is marked as \texttt{EDGE} in all fields & 0.687\\
& and strips, and thus it is deblended anyway. & \\
\texttt{INTERP} & The object contains one or more interpolated-over pixels. & 42.1\\
\texttt{MAYBE\_CR} & The object may be a cosmic ray. & 1.33\\
\texttt{MAYBE\_EGHOST} & The object may be a ghost produced by CCD electronics. & 0.143\\ \hline \hline

\multicolumn{2}{| l }{\emph{Image problem flags}: } \\ \cline{1-1}
\texttt{CANONICAL\_CENTER}* & The measurements use the center in the r-band rather than the local band. & 0 \\
\texttt{NOPROFILE}* & The object is either too small or too close to the edge and thus it is hard to estimate  & 0\\ 
& the radial flux profile. \\
\texttt{NOTCHECKED\_CENTER} & Similar to \texttt{NOTCHECKED}, but the affected pixels are close to object's center. & 0\\
\texttt{TOO\_LARGE} & The object is either too large to measure its profile or has a child greater than half of & 2.64E-6\\
&  the frame.  & \\
\texttt{BADSKY} & The local sky measurement is so bad and therefore the photometry is meaningless. & 0\\ \hline \hline

\multicolumn{2}{| l }{\emph{\AC{Suspicious object flags}}: } \\ \cline{1-1}
\texttt{DEBLENDED\_AS\_PSF} & \AC{If the deblending algorithm found this child is unresolved}. & 12.7 \\
\texttt{TOO\_FEW\_GOOD\_DETECTIONS} & \AC{This object doesn't have detections with good centroid in all bands.} & 38.7\\ 
\texttt{NO\_PETRO} & \AC{The code was not able to determine the Petrosian radius for this object.} & 26.4 \\ \hline \hline

\label{flags}
\end{tabular}}
\end{table*}

\AC{We now briefly discuss our strategy with respect to the flags listed in Table~\ref{flags}. First, for the flags listed in the \emph{Object status flags} section, we select objects that are \texttt{BINNED} but we do not use the \texttt{BRIGHT} flag. Since we originally selected objects from the \emph{Primary} catalog, which implies \texttt{!BRIGHT \&\& (!BLENDED || NODEBLEND || nchild == 0)}, we do not need to use the other flags listed in this section. Second, for the flags listed in the \emph{Raw data problem flags} section, we select objects that are not \texttt{SATURATED}. For the \texttt{EDGE} flag, we choose the object that has no \texttt{EDGE} flag, or is close to frame \texttt{EDGE} and has either \texttt{NODEBLEND} or \texttt{DEBLENDED\_AT\_EDGE} set. Third, for the flags listed in the \emph{Image problem flags} section, we select objects that are not \texttt{NOPROFILE}. Finally, for the flags listed in the \emph{Suspicious object flags} section, we exclude objects that have flag \texttt{DEBLENDED\_AS\_PSF} and contains either \texttt{TOO\_FEW\_GOOD\_DETECTION} or \texttt{NOPETRO}. The last step is essential because all objects it removes are suspicious objects, which we have visually examined.}

\AC{To summarize the previous discussion, we select all objects that are detected in \texttt{BINNED1}, are not flagged with either \texttt{SATURATED}, \texttt{NOPROFILE},  \texttt{DEBLEND\_PROBLEMS}, or \texttt{INTERP\_PROBLEMS}, and satisfy the \texttt{EDGE} criteria and removed the suspicious objects as discussed in the previous paragraph.  Thus we exclude objects with interpolation problems, but do not cut on \texttt{EDGE} since large galaxies can cross SDSS fields. Overall, these flag cuts keep $\sim145$ million sources, or 71.5\% of the data from the previous section.}

\subsection{Final sample selection}
\AC{Finally, following the discussion in Sections~\ref{ML} and~\ref{SG}, we select only those objects with dereddened r-band magnitudes between $17$ and $21$, and we exclude the objects with dereddened g- and i-band magnitudes fainter than 23. Our final cut is to choose galaxies by selecting those objects with $type\_r = 3$, which is the numerical value for galaxies. This produces a final galaxy catalog consisting of approximately $26$ million galaxies.}

\section{Full sample galaxy covariance matrix}\label{covtab}
In this Appendix, we present the full sample galaxy covariance matrix, as calculated from Equation~\ref{covmat} as described in Section~\ref{Correlation}. Given its size, we present a sample matrix for the full galaxy catalog in Tables~\ref{covmat1}, the full version and the covariance matrices for galaxy catalogs in four magnitude bins are available online in ASCII format.

\begin{table*}
\begin{center}
\caption{The sample covariance matrix for the full galaxy catalog, all values are in units of 10$^{-5}$. The full version and the covariance matrices for galaxy catalogs in four magnitude bins are available online in ASCII format.}
{\begin{tabular}{c c c c c c c c c c c c}\\ \hline \hline\label{covmat1}
Angle & 8.8101 & 6.8383 & 5.3078 & 4.1198 & 3.1978 & 2.4821 & 1.9265 & 1.4953 & 1.1607 & 0.9009 & \\ \hline
8.8101 & 0.0094 & 0.0078 & 0.0081 & 0.0088 & 0.0096 & 0.0103 & 0.0109 & 0.0114 & 0.0124 & 0.0138 & \\
6.8383 & 0.0078 & 0.0094 & 0.0107 & 0.0111 & 0.0114 & 0.0117 & 0.0122 & 0.0122 & 0.0129 & 0.0136 & \\
5.3078 & 0.0081 & 0.0107 & 0.0146 & 0.0155 & 0.0153 & 0.0157 & 0.0164 & 0.0161 & 0.0169 & 0.0181 & \\
4.1198 & 0.0088 & 0.0111 & 0.0155 & 0.0185 & 0.0188 & 0.0191 & 0.0198 & 0.0192 & 0.0201 & 0.0215 & \\
3.1978 & 0.0096 & 0.0114 & 0.0153 & 0.0188 & 0.0209 & 0.0223 & 0.0237 & 0.0236 & 0.0254 & 0.0272 & \\
2.4821 & 0.0103 & 0.0117 & 0.0157 & 0.0191 & 0.0223 & 0.0262 & 0.0286 & 0.0291 & 0.0316 & 0.0342 & \\
1.9265 & 0.0109 & 0.0122 & 0.0164 & 0.0198 & 0.0237 & 0.0286 & 0.0322 & 0.0335 & 0.0367 & 0.0397 & \\
1.4953 & 0.0114 & 0.0122 & 0.0161 & 0.0192 & 0.0236 & 0.0291 & 0.0335 & 0.0368 & 0.0411 & 0.0448 & \\
1.1607 & 0.0124 & 0.0129 & 0.0169 & 0.0201 & 0.0254 & 0.0316 & 0.0367 & 0.0411 & 0.0472 & 0.0521 & \\
0.9009 & 0.0138 & 0.0136 & 0.0181 & 0.0215 & 0.0272 & 0.0342 & 0.0397 & 0.0448 & 0.0521 & 0.0588 & \\ \hline
\end{tabular}}
\end{center}
\end{table*}

\end{document}